\documentclass[sigplan,10pt]{acmart}
\renewcommand\footnotetextcopyrightpermission[1]{}
\usepackage{epsfig,endnotes}

\def\BibTeX{{\rm B\kern-.05em{\sc i\kern-.025em b}\kern-.08emT\kern-.1667em\lower.7ex\hbox{E}\kern-.125emX}}

\usepackage{color}
\usepackage{multirow}
\usepackage{subcaption}
\usepackage{xspace}
\usepackage[title]{appendix}
\usepackage{multirow}
\usepackage[inline]{enumitem}
\usepackage{siunitx}
\DeclareSIUnit{\microsecond}{\SIUnitSymbolMicro s}

\usepackage[utf8]{inputenc}
\usepackage{pgfplots}
\usetikzlibrary{patterns}

\DeclareMathAlphabet\mathbfcal{OMS}{cmsy}{b}{n}

\hypersetup{
  colorlinks=true,      % false: boxed links; true: colored links
  linkcolor=violet,       % color of internal links
  citecolor=violet,    % color of links to bibliography
  filecolor=cyan,       % color of file links
  urlcolor=violet          % color of external links
}

\newenvironment{denseitemize}{
\begin{itemize}[topsep=2.5pt, partopsep=0pt, leftmargin=1.5em]
  \setlength{\itemsep}{2.5pt}
  \setlength{\parskip}{0pt}
  \setlength{\parsep}{0pt}
}{\end{itemize}}

\newenvironment{denseenum}{
\begin{enumerate}[topsep=2.5pt, partopsep=0pt, leftmargin=1.5em]
  \setlength{\itemsep}{2.5pt}
  \setlength{\parskip}{0pt}
  \setlength{\parsep}{0pt}
}{\end{enumerate}}

\usepackage{enumitem}
\setitemize{noitemsep,topsep=0pt,parsep=0pt,partopsep=0pt}

\DeclareMathAlphabet{\mathcal}{OMS}{cmsy}{m}{n}

\newcommand{\jack}[1]{{\color{purple}[Jack: #1]}}
\newcommand{\rashmi}[1]{{\color{blue}[Rashmi says: #1]}}
\newcommand{\shivaram}[1]{{\color{green}[Shivaram: #1]}}

\renewcommand{\jack}[1]{}
\renewcommand{\rashmi}[1]{}
\renewcommand{\shivaram}[1]{}

\newcommand{\codedCompShortlist}{\cite{lee2018speeding,yu2017polynomial,dutta2017coded,konstantinidis2018leveraging,konstantinidis2018leveraging,tandon2017gradient,karakus2017straggler}\xspace}

\newcommand{\handCraftedCodedComputeRefs}{\cite{lee2018speeding,dutta2016short,li2016unified,yu2017polynomial,wang2018sparse,dutta2017coded,reisizadeh2017coded,mallick2018rateless,yu2018lagrange}\xspace}

\newcommand{\etal}{et al.\xspace}

\newcommand{\func}{\mathcal{F}\xspace}
\newcommand{\funcBold}{\mathbfcal{F}\xspace}
\newcommand{\funcRed}{\func_P\xspace}
\newcommand{\enc}{\mathcal{E}\xspace}
\newcommand{\dec}{\mathcal{D}\xspace}
\newcommand{\data}{X\xspace}
\newcommand{\dataOne}{X_{1}\xspace}
\newcommand{\dataTwo}{X_{2}\xspace}
\newcommand{\dataThree}{X_{3}\xspace}

\newcommand{\dataK}{X_{k}\xspace}
\newcommand{\outOne}{\func(\dataOne)\xspace}
\newcommand{\outTwo}{\func(\dataTwo)\xspace}
\newcommand{\outThree}{\func(\dataThree)\xspace}

\newcommand{\outK}{\func(\dataK)\xspace}

\newcommand{\outBold}{\funcBold(\mathbf{\data})\xspace}
\newcommand{\parity}{P\xspace}

\newcommand{\outP}{\func(\parity)\xspace}
\newcommand{\outPBold}{\funcBold(\mathbf{\parity})\xspace}
\newcommand{\outRed}{\funcRed(\parity)\xspace}

\newcommand{\query}{query\xspace}
\newcommand{\queries}{queries\xspace}
\newcommand{\Query}{Query\xspace}
\newcommand{\Queries}{Queries\xspace}
\newcommand{\prediction}{prediction\xspace}
\newcommand{\predictions}{predictions\xspace}

\newcommand{\Predictions}{Predictions\xspace}

\newcommand{\slaFull}{service level objective\xspace}
\newcommand{\sla}{SLO\xspace}

\newcommand{\tech}{ParM\xspace}

\newcommand{\paritymodel}{parity model\xspace}
\newcommand{\paritymodels}{parity models\xspace}
\newcommand{\Paritymodel}{Parity model\xspace}

\newcommand{\modelinstance}{model instance\xspace}
\newcommand{\modelinstances}{model instances\xspace}

\newcommand{\Modelinstances}{Model instances\xspace}

\newcommand{\basemodel}{deployed model\xspace}
\newcommand{\basemodels}{deployed models\xspace}

\newcommand{\encdec}{encoder and decoder\xspace}
\newcommand{\encdecs}{encoders and decoders\xspace}
\newcommand{\Encdec}{Encoder and decoder\xspace}
\newcommand{\encoder}{encoder\xspace}

\newcommand{\decoder}{decoder\xspace}

\newcommand{\codinggroup}{coding group\xspace}

\newcommand{\parityBatch}{parity batch\xspace}

\newcommand{\modelserver}{prediction serving system\xspace}
\newcommand{\modelservers}{prediction serving systems\xspace}

\newcommand{\Modelservers}{Prediction serving systems\xspace}

\newcommand{\modelservices}{prediction services\xspace}

\newcommand{\modelserving}{prediction-serving\xspace}

\newcommand{\baseEqual}{Equal-Resources\xspace}
\newcommand{\baseEqualShort}{E.R.\xspace}

\newcommand{\nn}{neural network\xspace}
\newcommand{\nns}{neural networks\xspace}
\newcommand{\Nn}{Neural network\xspace}

\newcommand{\Section}{\S}
\newcommand{\highlightTailToMed}{3.5\xspace}
\newcommand{\highlightPercentDiff}{48\xspace}

\usepackage{fancyhdr}
\fancypagestyle{firststyle}
{
   \fancyhf{}
   \chead{\footnotesize This paper is superseded by the ACM SOSP 2019 paper ``Parity Models: Erasure-Coded Resilience for Prediction Serving Systems''}
}

\begin{document}

%don't want date printed
\date{}

\title{Parity Models: A General Framework for Coding-Based Resilience in ML Inference}

\renewcommand{\shortauthors}{}
\renewcommand{\shorttitle}{}

\author{Jack Kosaian}
\affiliation{%
  \institution{Carnegie Mellon University}
}
\email{jkosaian@cs.cmu.edu}
\author{K.~V. Rashmi}
\affiliation{%
  \institution{Carnegie Mellon University}
}
\email{rvinayak@cs.cmu.edu}
\author{Shivaram Venkataraman}
\affiliation{%
  \institution{University of Wisconsin-Madison}
}
\email{shivaram@cs.wisc.edu}

\settopmatter{printfolios=true,printacmref=false}
\maketitle

% Use the following at camera-ready time to suppress page numbers.
% Comment it out when you first submit the paper for review.
%\thispagestyle{empty}
% \pagestyle{fancy}
% \fancyhead[CO,CE]{DRAFT: DO NOT DISTRUBUTE}
% \fancyfoot[CO,CE]{DRAFT: DO NOT DISTRUBUTE}
% \fancyfoot[LE,RO]{\thepage}
% \fancypagestyle{plain}{\pagestyle{fancy}}

\thispagestyle{firststyle}
\section*{Abstract}
Machine learning models are becoming the primary work-horses for many applications. Production services deploy models through prediction serving systems that take in que-ries and return predictions by performing inference on machine learning models. In order to scale to high query rates, prediction serving systems are run on many machines in cluster settings, and thus are prone to slowdowns and failures that inflate tail latency and cause violations of strict latency targets. Current approaches to reducing tail latency are inadequate for the latency targets of prediction serving, incur high resource overhead, or are inapplicable to the computations performed during inference.

We present ParM, a novel, general framework for making use of ideas from erasure coding and machine learning to achieve low-latency, resource-efficient resilience to slowdowns and failures in prediction-serving systems. ParM \emph{encodes multiple queries} together into a single \emph{parity query} and performs inference on the parity query using a \emph{parity model}. A \emph{decoder} uses the output of a parity model to reconstruct approximations of unavailable predictions. ParM uses neural networks to \emph{learn} parity models that enable simple, fast encoders and decoders to reconstruct unavailable predictions for a \textit{variety of inference tasks such as image classification, speech recognition, and object localization}. We build ParM atop an open-source prediction-serving system and through extensive evaluation show that ParM improves overall accuracy in the face of unavailability with low latency while using 2-4$\times$ less additional resources than replication-based approaches. ParM reduces the gap between 99.9th percentile and median latency by up to $3.5\times$ compared to approaches that use an equal amount of resources, while maintaining the same median.

\section{Introduction} \label{sec:intro}
Machine learning has become ubiquitous in production services~\cite{agarwal2014laser,baylor2017tfx} and user-facing applications~\cite{siri, alexa, google-lens}. This has increased the importance of inference, the process of returning a prediction from a trained machine learning model. \textit{\Modelservers} are platforms that host machine learning models for inference and deliver model \predictions for input \queries. Numerous \modelservers are being developed both by cloud service providers~\cite{aws-ml,azure-ml-studio,google-ai} as well as by open-source communities~\cite{crankshaw2017clipper,olston2017tensorflow,mxnet-model-server,mmlspark}.

\begin{figure}[t]
    \centering
    \includegraphics[width=0.8\linewidth]{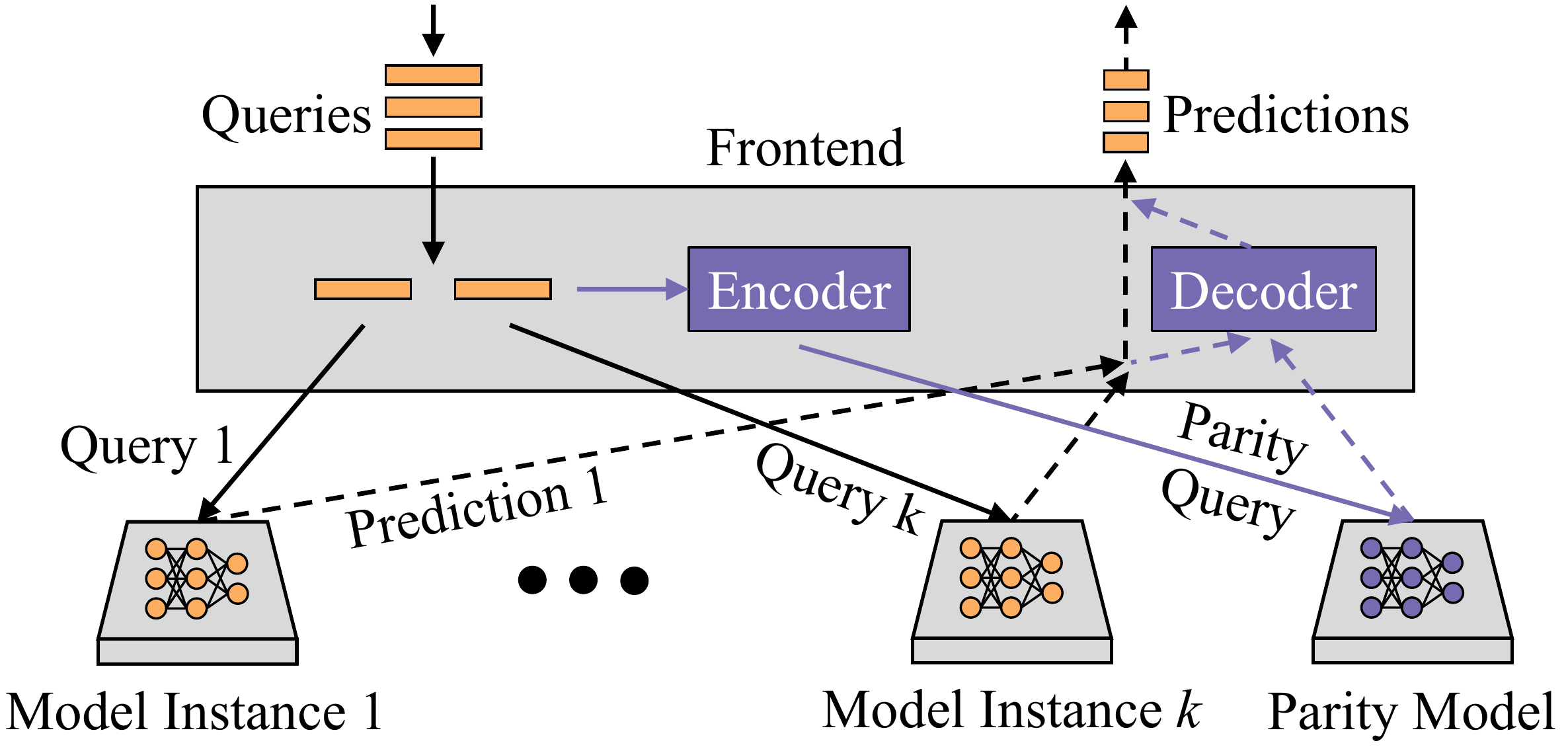}
    \caption{Architecture of a \modelserver along with components introduced by \tech (darkly shaded).}
    \label{fig:redundancy-modes:og}
    \vspace{-0.2in}
\end{figure}

In order to meet the demands of user-facing production services, \modelservers must deliver \predictions with low latency (e.g., within tens of milliseconds~\cite{crankshaw2017clipper}). Similar to other latency-sensitive services, \modelservices must adhere to strict \slaFull{s} (\sla{s}). \Queries that are not completed by their \sla are often useless to applications~\cite{agarwal2014laser}.
In order to reduce \sla violations, \modelservers must \textit{minimize tail latency}.

\Modelservers often employ distributed architectures to support the high throughput required by production services~\cite{crankshaw2017clipper,lee2018towards}.  As depicted in Figure~\ref{fig:redundancy-modes:og} (ignoring the dark components for the moment), a \modelserver consists of a frontend which receives \queries and dispatches them to one or more \modelinstances. \Modelinstances perform inference and return \predictions. This distributed setup is typically run in large-scale, multi-tenant clusters (e.g., public clouds), where tail latency inflation is a common problem~\cite{dean2013tail}. There are numerous causes of inflated tail latencies in these settings, such as multi-tenancy and resource contention~\cite{xu2013bobtail,hauswald2015djinn,iorgulescu2018perfiso}, hardware unreliability and failures~\cite{ananthanarayanan2010reining}, and other complex runtime interactions~\cite{ananthanarayanan2013efficient}. Within the context of \modelservers, network and computation contention have both been shown as potential bottlenecks~\cite{hauswald2015djinn,crankshaw2017clipper}, and routines like loading a new model can also cause latency spikes~\cite{olston2017tensorflow}.

Due to the many causes of tail latency inflation, it is important for mitigations to be agnostic to the cause of slowdown~\cite{dean2013tail}. However, current agnostic approaches for mitigating tail latency inflation are either inadequate for the low latency typical of \modelserving~\cite{zaharia2008improving,yadwadkar2014wrangler} or replicate queries, requiring significant resource overhead~\cite{ananthanarayanan2012why,ananthanarayanan2013efficient,vulimiri2012more}.

Erasure codes are popular tools employed for imparting resilience to data unavailability while remaining agnostic to the cause of unavailability and using less resources than replication-based approaches. Erasure codes are used in various settings such as storage~\cite{patterson1988raid,facebookECsavings2010_forACM,huang2012erasure,rashmi2014hitchhiker,rashmi2016eccache,yan2017tiny} and communication~\cite{rizzo1997effective,richardson2008modern}. An erasure code encodes $k$ data units to produce $r$ redundant units called ``parities'' in such a way that any $k$ of the total $(k+r)$ data and parity units are sufficient for a \decoder to recover the original $k$ data units. The overhead incurred by an erasure code is $\frac{k+r}{k}$, which is typically much less than that of replication (by setting $r < k$).

\pagestyle{plain}
A number of recent works have explored using erasure codes for alleviating the effects of slowdowns and failures that occur in distributed \textit{computation}~\codedCompShortlist. In this setup, called ``coded-computation,'' erasure coding is used for recovering the \textit{outputs} of a deployed computation over data units. In coded-computation, data units are encoded into parity units, and the deployed computation is performed over all data and parity units in parallel. A decoder then uses the outputs from the fastest $k$ of these computations to reconstruct the outputs corresponding to the original data units. For a \modelserver, employing coded-computation would involve encoding \textit{\queries} such that a \decoder can recover unavailable \textit{\predictions} from slow or failed \modelinstances.

The primary differences between coded-computation and the traditional use of erasure codes in storage and communication come from (1) performing computation over encoded data and (2) the need for an erasure code to recover the results of computation over data units rather than the data units themselves. Whereas traditional applications of erasure codes involve encoding data units and decoding from a subset of data and parity units, in coded-computation one decodes by using the \textit{output of computation over data and parity units}. This difference calls for fundamentally rethinking the design of erasure codes, as many of the erasure codes which have found widespread use in storage and communication (e.g., Reed-Solomon codes~\cite{reed1960polynomial}), are applicable only to a highly restricted class of computations~\cite{lee2018speeding}. 

As erasure codes can correct slowdowns with low latency and require less resource overhead than replication-based techniques, enabling the use of coded-computation in \modelservers has promising potential for efficient mitigation of tail latency inflation. However, the complex non-linear components common to popular machine learning models deployed in \modelservers, like \nns, make it challenging to design effective coded-computation solutions. 

Most prior coded-computation techniques support only rudimentary computations such as linear functions, low-degree polynomials, and a subset of matrix operations~\handCraftedCodedComputeRefs. These prior approaches are unable to support the complex non-linear computations common to popular machine learning models like neural networks, making them inadequate for machine learning inference. 

Kosaian \etal\cite{kosaian2018learning} proposed the first approach enabling coded-computation for machine learning inference by introducing two key ideas: (1) \textit{Allowing for approximation}: Prior coded-computation approaches focused on recovering unavailable outputs \textit{exactly}. Kosaian \etal\cite{kosaian2018learning} introduced the notion of approximation in coded-computation by observing that the exact reconstruction requirement can be relaxed for machine learning inference because the \predictions from machine learning models are themselves approximate. (2) \textit{Using learning:} Prior coded-computation approaches focused on hand-designing encoders and decoders. Kosaian \etal\cite{kosaian2018learning} proposed the first learning-based approach for coded-computation by designing encoders and decoders as machine learning models and learning an erasure code for imparting resilience over a given computation.
Using this approach, Kosaian \etal\cite{kosaian2018learning} learn encoders and decoders for imparting resilience over \nn inference. This marked a significant step forward from prior coded-computation techniques, which, at the time, were unable to support even simple non-linear functions.

While the learning-based approach proposed by Kosaian \etal\cite{kosaian2018learning} showcases the promise for imparting coding-based resilience to neural network inference, the approach introduces a number of challenges: (1) \textit{High latency of reconstructions:} The learned encoders and decoders proposed by Kosaian \etal\cite{kosaian2018learning} are computationally expensive. As will be described in \Section\ref{sec:learning}, these encoders and decoders can be up to $7\times$ slower than the deployed models over which they are intended to impart resilience. This high latency makes this approach appropriate for reducing only the far end of tail latency. (2) \textit{Need for hardware acceleration:} The latency of learned encoders and decoders can potentially be reduced through the use of hardware acclerators (e.g., GPU, TPU). However, doing so requires the use of a more expensive \modelserver frontend which, as will be described in \Section\ref{sec:design}, is the ideal location for placing encoders and decoders.

To address these challenges, we present a \textit{fundamentally new, general framework for coded-computation}, aimed at mitigating tail latency inflation in \modelservers, which we call \tech (\textbf{par}ity \textbf{m}odels). 
Unlike conventional coded-computation approaches, which design new erasure codes, \tech allows for 
simple, fast encoders and decoders and instead employs specialized units for performing computation over parities, which we call \textit{parity models}, as depicted in Figure~\ref{fig:redundancy-modes:og}.
\tech encodes \queries together into a parity \query. A \paritymodel transforms the parity query such that its output enables the \decoder to reconstruct unavailable \predictions. 
\tech designs \paritymodels as \nns and learns a transformation over parity \queries that enables simple encoders and decoders (such as addition and subtraction) to impart resilience for a given deployed model.

The contributions of this paper are as follows.
\begin{denseenum}
\item We present \tech, a novel, general framework for efficient use of coded-computation in \modelservers.
\item We propose and design \textit{parity models} as a new building block in coded computation. Our approach of using parity models makes \tech applicable to a wide \textit{variety of inference tasks such as image classification, speech recognition, and object localization}.
\item We have built \tech atop  Clipper~\cite{crankshaw2017clipper}, a popular open-source \modelserver.
\item We extensively evaluate \tech's ability to reduce tail latency and improve accuracy under unavailability. Our evaluations show that \tech significantly reduces tail latency and helps in improving overall accuracy in the face of unavailablility, while using 2-4$\times$ less additional resources than replication-based approaches.
\item \textit{Accuracy}: \tech accurately reconstructs unavailable \predictions for a  variety of inference tasks such as image classification, speech recognition, and object localization, and for a variety of \nns. For example, using only half of the additional resources as replication, \tech's reconstructions from ResNet-18 models on various tasks are up to 89\% more accurate than approaches that return default \predictions in the face of unavailability. Further, \tech can reconstruct unavailable \predictions to be within a 6.5\% difference in accuracy compared to if the original \predictions were not slow or failed.
\item \textit{Latency}: \tech reduces tail latency across a variety of \query rates, levels of background load, and amounts of redundancy. For example, \tech reduces 99.9th percentile latency in the presence of load imbalance for a ResNet-18 model by up to $\highlightPercentDiff\%$ compared to a baseline that uses the same amount of resources as \tech, while maintaining the same median. This brings tail latency up to $\highlightTailToMed\times$ closer to median latency, enabling \tech to return predictions with predictable latencies in the face of unavailability.
\end{denseenum}

Our results show the promise of \tech's approach of using parity models as building blocks in coded-computation for machine learning inference. This framework opens new doors for imparting resource-efficient resilience to \modelservers.

\pagestyle{plain}
\section{Background and Motivation}\label{sec:background}
This section describes the architecture of \modelservers, as well as the challenges and opportunities for improvement. 
This discussion is informed by popular \modelservers~\cite{crankshaw2017clipper,olston2017tensorflow} and conversations with service providers via personal communication.

We also describe current approaches for imparting resilience to distributed computation, as well as their limitations, which we specifically address in this paper.

\subsection{\Modelservers}
A \modelserver hosts machine learning models for inference; it accepts \queries from clients, performs inference on hosted models, and returns \predictions. We refer to a model hosted for inference as a ``\basemodel.'' 

As depicted in Figure~\ref{fig:redundancy-modes:og} (ignoring the dark components), \modelservers have two types of components: a frontend and \modelinstances. The frontend receives \queries and dispatches them to \modelinstances for inference.\footnote{Not all \modelserving frameworks (e.g., TensorFlow Serving~\cite{olston2017tensorflow}) have a specific frontend process. These systems make use of a load balancer to distribute \queries, which acts in many ways like the frontend we describe.} \Modelinstances are containers or processes that contain a copy of the \basemodel and return \predictions by performing inference on the \basemodel.

% \noindent
\textbf{Scale-out architecture.} \Modelservers use scale-out architectures to serve \predictions with low latency and high throughput and to overcome the memory and processing limitations of a single server~\cite{lee2018towards}. In such a setup, multiple \modelinstances are deployed on separate servers, each containing a copy of the same \basemodel~\cite{crankshaw2017clipper}. The frontend distributes \queries to \modelinstances according to a load-balancing strategy (e.g., single-queue, round-robin).

% \noindent
\textbf{Inference hardware and \query batching.} \Modelservers use a variety of hardware for performing inference, including GPUs~\cite{crankshaw2017clipper}, CPUs~\cite{hazelwood2018applied,park2018deep,zhang2018deepcpu}, TPUs~\cite{jouppi2017datacenter}, and FPGAs~\cite{chung2018serving}. As some hardware is optimized for batched operation (e.g., GPUs), some \modelservers will buffer and batch \queries at the frontend and dispatch \queries to \modelinstances in batches~\cite{crankshaw2017clipper,olston2017tensorflow}. However, as batching induces latency, many systems perform minimal or no batching~\cite{chung2018serving,zhang2018deepcpu}, especially when using hardware that is not tailored for batched operation (e.g., FPGAs~\cite{chung2018serving}, CPUs~\cite{hazelwood2018applied}).

\subsection{Challenges and opportunity}
\label{sec:background:bottlenecks}
As described above, \modelservers are often run in a distributed fashion and make use of many cluster resources (e.g., compute, network). These systems are thus prone to the slowdowns and failures common to cloud and cluster settings. Left unmitigated, these slowdowns inflate tail latency. \Modelservers must therefore employ some means to mitigate the effects of slowdowns in order to meet latency \sla{s}. Due to the many causes of slowdowns, such as those described in \Section\ref{sec:intro}, it is important for mitigations to be agnostic to the cause of slowdowns~\cite{dean2013tail}. However, as we describe next, existing agnostic techniques are either inadequate for the low latency required of \modelserving, or rely on resource-intensive replication.

Speculative execution techniques~\cite{mapreduce-speculative-execution, zaharia2008improving} wait for a task to progress before detecting that the task is slow and taking corrective action. Other techniques predict when slowdowns will occur, but incur scheduling delays in mitigating slowdowns~\cite{yadwadkar2014wrangler}. While the delays incurred by these approaches are negligible for the timescales of data analytics tasks, they are inadequate for mitigating slowdowns in \modelserving, where \queries are expected to be processed within tens to hundreds of milliseconds~\cite{crankshaw2017clipper,gujarati2017swayam}.

Replication-based techniques~\cite{ananthanarayanan2012why,ananthanarayanan2013efficient,vulimiri2012more,dean2013tail} mitigate slowdowns \textit{proactively} by sending duplicate \queries to replicas of an underlying task that utilize separate resources (e.g., \basemodels on separate servers) and waiting only for the first replica to respond. Thus, a system which replicates a \query $k$ times can tolerate $(k-1)$ slow or failed responses. By proactively issuing redundant \queries at the same time as the original \query, replication mitigates slowdowns without additional delay for detecting slowdowns. However, replication requires high resource overhead, as replicating \queries $k$ times requires $k$-times as many resources to handle increased load. 
Attempting to reduce this overhead by retrying \queries only if a response has not been received by a certain time (i.e., ``hedged requests''~\cite{dean2013tail}) results in reducing only the far end of tail latency due to delays induced by waiting, similar to the speculative techniques described above.

Erasure codes are used in  storage~\cite{patterson1988raid,facebookECsavings2010_forACM,rashmi2014hitchhiker,rashmi2016eccache,yan2017tiny} and communication~\cite{rizzo1997effective,richardson2008modern} to mitigate slowdowns and failures both with low latency and with less resources than replication. Leveraging ideas from erasure codes to recover unavailable outputs from inference---rather than recovering unavailable data as in traditional applications---could potentially alleviate slowdowns and failures in a resource-efficient manner in \modelservers while remaining agnostic to the cause of unavailability.

\begin{figure}[tp]
	\centering
	\includegraphics[width=0.7\linewidth]{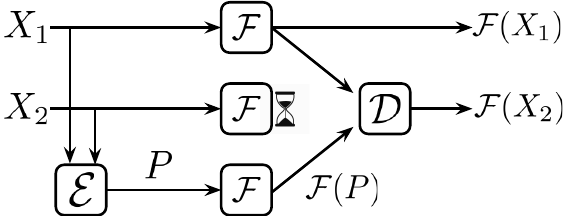}
    \caption{Abstract example of coded-computation with $k=2$ original units and $r=1$ parity units.}
    \label{fig:background:cc}
\end{figure}

\subsection{Coded-computation and its challenges}\label{sec:background:ec}
The approach of using erasure codes for alleviating the effects of slowdowns and failures in \emph{computation} is termed ``coded-computation.'' Coded-computation differs fundamentally from the traditional use of erasure codes. Erasure codes have traditionally been used for recovering unavailable data units using a subset of data and parity units. In contrast, under coded-computation, (1) computation is performed over encoded data and (2) the goal is to recover unavailable \emph{outputs of computation over data units} using a subset of the outputs of computation over data and parity units.

% \noindent
\textbf{Example.} Consider an example in Figure~\ref{fig:background:cc}. Let $\func$ be a computation that is deployed on two servers. Let $\dataOne$ and $\dataTwo$ be inputs to the computation. The goal is to return $\outOne$ and $\outTwo$. In a \modelserver, $\func$ is a \basemodel and $\dataOne$ and $\dataTwo$ are \queries. Coded-computation adds an \encoder $\enc$ and a \decoder $\dec$, along with a third copy of $\func$ for tolerating one of the copies of $\func$ being unavailable. The \encoder produces a parity $\parity = \enc(\dataOne, \dataTwo)$. The parity is dispatched to the third copy of $\func$ to produce $\outP$. Given $\outP$ and any one of $\{\outOne,\outTwo\}$, the \decoder reconstructs the unavailable output. In the example in Figure~\ref{fig:background:cc}, the second computation is slow. The \decoder produces a reconstruction of $\outTwo$ as $\dec(\outOne, \outP)$.

% \noindent
\textbf{General parameters.} More generally, given $k$ instances of $\func$, 
for $k$ queries $\dataOne,\ldots,\dataK$, the goal is to output $\outOne,\ldots,$ $\outK$. 
To tolerate any $r$ of these being unavailable, the encoder generates $r$ parity queries that are operated on by $r$ redundant copies of $\func$. The decoder acts on any $k$ outputs of these $(k+r)$ instances of $\func$ to recover $\outOne, \ldots, \outK$.

\begin{table}[t]
	\renewcommand{\arraystretch}{1.4}
	\centering
	\begin{tabular}[t]{ccc}
    	\toprule
		$\outBold$ & $\outPBold$ & \textbf{Desired $\outPBold$} \\ 
		\midrule
        $2X$ & $2\dataOne + 2\dataTwo$ & $2\dataOne + 2\dataTwo$ \\
        									   $X^2$ & $\dataOne^2 + 2\dataOne\dataTwo + \dataTwo^2$ & $\dataOne^2 + \dataTwo^2$ \\ 
	\end{tabular}
    \caption{Toy example with parity $\parity = \dataOne + \dataTwo$ showing the challenges of coded-computation on non-linear functions.}
    \label{table:background:cc}
\end{table}

% \noindent
\textbf{Challenges.} Coded-computation is straightforward when the underlying computation $\func$ is a \textit{linear} function.   
A function $\func$ is linear if, for any inputs $\dataOne$ and $\dataTwo$, and any constant $a$: (1) $\func(\dataOne + \dataTwo) = \outOne + \outTwo$ and (2) $\func(a\dataOne) = a\outOne$. Many of the erasure codes used in traditional applications, such as Reed-Solomon codes in storage, can recover from unavailability of  any linear function~\cite{lee2018speeding}. 
For example, consider having $k=2,\ r=1$. Suppose $\func$ is a linear function as in the first row of Table~\ref{table:background:cc}. Here, even a simple parity $\parity=\dataOne + \dataTwo$ suffices since  $\func(P)=\outOne + \outTwo$ and the decoder can subtract the available output from the parity output to recover the unavailable output. The same argument holds for any linear $\func$. However, a non-linear $\func$ significantly complicates the scenario. For example, consider $\func$ to be the simple non-linear function in the second row of Table~\ref{table:background:cc}. As shown in the table, $\func(P) \neq \outOne + \outTwo$, and even for this simple function, $\func(P)$ involves complex non-linear interactions of the inputs which makes decoding difficult.

Handling non-linear computation is key to using coded-computation in \modelservers due to the many non-linear components of popular machine learning models, such as \nns. While \nns do contain linear components (e.g., matrix multiplication), they also contain many non-linear components (e.g., activation functions, max-pooling), which make the overall function computed by a \nn non-linear.

As discussed in \Section\ref{sec:intro}, most prior techniques approach coded-computation by hand-crafting new encoders and decoders.
However, due to the challenge of handling non-linear computations, these approaches support only rudimentary computations~\handCraftedCodedComputeRefs, and hence are unable to support popular machine learning models like \nns. Kosaian \etal\cite{kosaian2018learning} present the first coded-computation approach applicable for machine learning inference. We discuss benefits and challenges of this approach below.

\subsection{Benefits and challenges of learning a code}\label{sec:learning}
As illustrated in \Section\ref{sec:background:ec}, it is challenging to hand-craft erasure codes for the non-linear components common to \nns. This problem is further complicated by the multitude of mathematical components employed in \nns (e.g., types of layers, activation functions); even if one developed an erasure code suitable for one \nn, the approach might not work for other \nns.

To overcome this, Kosaian \etal\cite{kosaian2018learning} observe that erasure codes for coded-computation can be \textit{learned}. Using machine learning models for encoders and decoders, designing an erasure code simply involves training \encoder and \decoder models. Consider again the example in Figure~\ref{fig:background:cc}. An optimization problem for learning \encoder $\enc$ and \decoder $\dec$ for this example is: given $\func$, train $\enc$ and $\dec$ so as to minimize the difference between $\widehat{\outTwo}$, the output of the decoder, and $\outTwo$, for all pairs $(\dataOne, \dataTwo)$, and with $\parity = \enc(\dataOne, \dataTwo)$ and $\widehat{\outTwo} = \dec(\outOne, \outP)$. One distinction of using learned \encoder{s} and \decoder{s} as opposed to traditional hand-crafted ones is that reconstructions of unavailable outputs will be \textit{approximations} of the function outputs that would be returned if they were not slow or failed. This is appropriate for \modelservers because the \predictions returned by \basemodels themselves are approximations. Further, any decrease in accuracy due to reconstruction is only incurred in the case when a model is slow to return a \prediction. In this scenario, \modelservices prefer to return an approximate \prediction rather than a late one~\cite{agarwal2014laser}.

Using this approach, Kosaian \etal\cite{kosaian2018learning} designed \nn encoders and decoders for imparting resilience over \nn inference. For example, the approach enables unavailable predictions from a ResNet-18 model trained for CIFAR-10 to be reconstructed with up to 82\% accuracy. This is a
small drop in accuracy compared to the accuracy of the deployed model (93\%), and the drop only occurs when the deployed model is unavailable. This marked a significant step forward from prior coded-computation techniques, which, at the time, were unable to support even simple non-linear functions.

% \noindent
\textbf{Challenges of learned codes.} While the approach taken by Kosaian \etal\cite{kosaian2018learning} showcases the potential of using machine learning for coded-computation, it also reveals a challenge: neural network \encoder{s} and \decoder{s} add significant latency to reconstruction. Recall that a coded-computation technique can reconstruct unavailable outputs only after (1) encoding, (2) performing $k$ out of $(k+r)$ computations, and (3) decoding. As \nns are computationally expensive, encoding and decoding with \nns adds significant latency to this process, and thus limits opportunities for tail latency reduction. Indeed, we found that the average latency of encoding and decoding using convolutional \nns~\cite{kosaian2018learning} was up to $7\times$ higher than that of the \basemodel, making this approach appropriate only for reducing the far end of tail latency. While the latency of learned encoders and decoders can potentially be reduced through the use of hardware accelerators, this necessitates a beefier, expensive frontend which, as will be described in \Section\ref{sec:design:design}, is the ideal location for encoders and decoders.
\section{Parity Models Framework}\label{sec:design}
\begin{figure}[tp]
	\centering
	\includegraphics[width=0.7\linewidth]{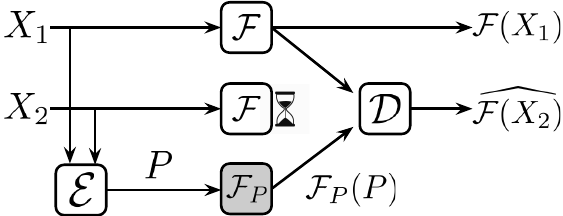}
    \caption{Abstract example of coded-computation when using a parity model ($\funcRed$) with $k=2$ original units and $r=1$ parity units.}
    \label{fig:design:cc_pm}
\end{figure}

In order to overcome both the challenge of performing coded-computation over non-linear functions as well as the high latency of learned encoders and decoders, we take a \textit{fundamentally new approach to coded-computation} in \tech. Rather than designing new \encdecs, \tech uses simple, fast \encdecs and instead designs a new \textit{computation over parities}, called a ``\paritymodel.'' 
As depicted in Figure~\ref{fig:design:cc_pm}, instead of the extra copy of $\func$ deployed by current coded-computation approaches, \tech introduces a \paritymodel, which we denote as $\funcRed$. 
The key challenge of this approach is to design a \paritymodel that enables reconstruction for a given computation $\func$. \tech addresses this by designing \paritymodels as \nns, and \emph{learning} a \paritymodel that enables a simple encoder and decoder to reconstruct slow or failed \predictions.

By learning a \paritymodel and using simple, fast \encdecs, \tech is (1) able to impart resilience to modern machine learning models, like \nns, while (2) operating with low latency without requiring expensive hardware acceleration for encoding and decoding.

% \noindent
\textbf{Setting and notation.} We first describe \tech in detail for imparting resilience to any one out of $k$ \predictions experiencing slowdown or failure (i.e., $r=1$). This setting is motivated by measurements of production clusters~\cite{rashmi2014hitchhiker,rashmi2016eccache}. Section~\ref{sec:design:concurrent} describes how the proposed approach can tolerate multiple unavailabilities (i.e., $r>1$) as well. We will continue to use the notation of $\func$ to represent the \basemodel, $\data_i$ to represent a \query, $\func(\data_i)$ to represent a \prediction resulting from inference on $\func$ with $\data_i$. We will let $\widehat{\func(\data_i)}$ represent a reconstruction of $\func(\data_i)$ when $\func(\data_i)$ is unavailable.

\begin{figure}[t]
    \centering
    \includegraphics[width=0.9\linewidth]{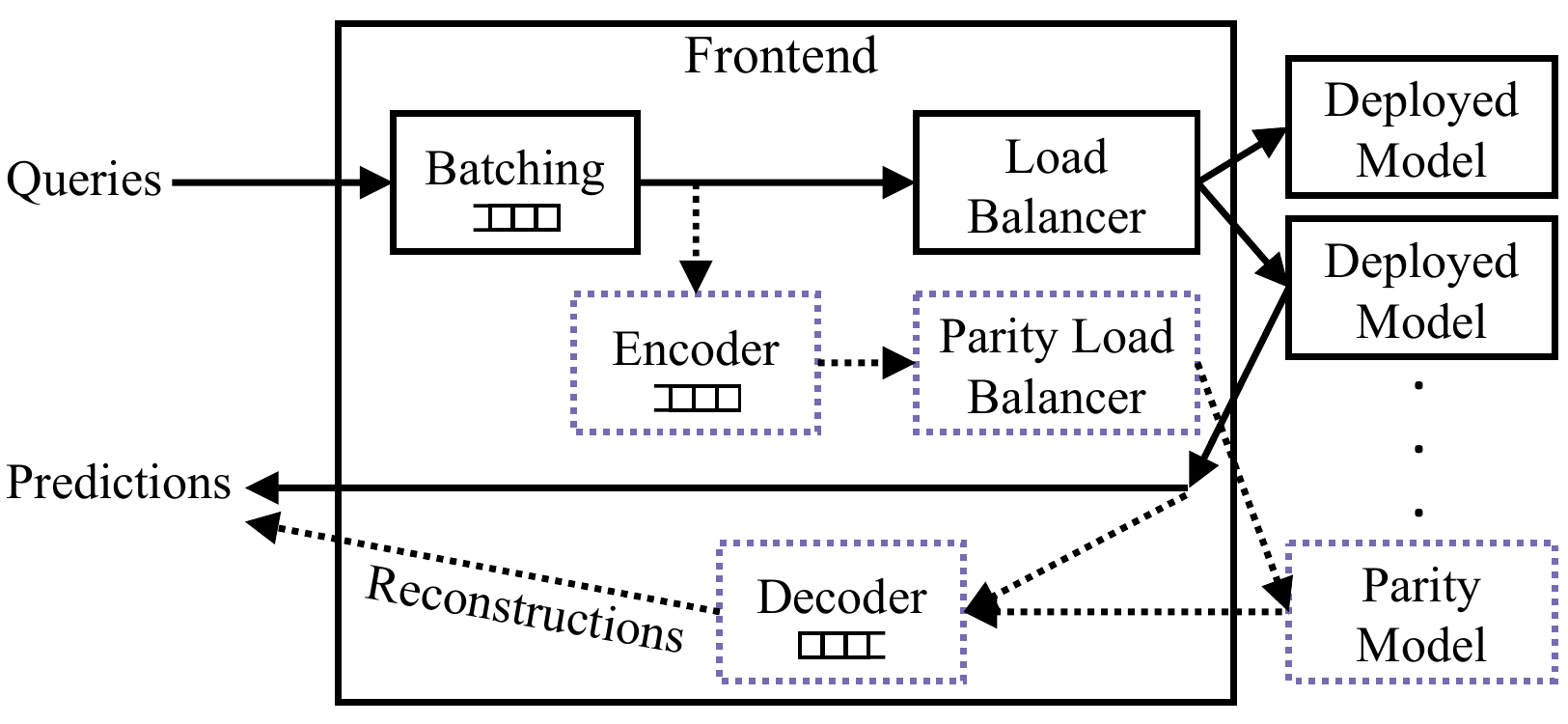}
    \caption{Components of a \modelserver and those added by \tech (dotted). Queues indicate components which may group \queries/\predictions (e.g., \codinggroup).}
    \label{fig:design:internal}
\end{figure}

\subsection{System architecture}\label{sec:design:design}
The architecture of \tech is shown in Figure~\ref{fig:design:internal}. \tech builds atop a typical \modelserver architecture that has $m$ instances of a \basemodel. 
\Queries sent to the frontend are batched (according to a batching policy) and dispatched to a \modelinstance for inference on the \basemodel. \Query batches\footnote{We use the terms ``batch'' and ``\query batch'' to refer to one or more \queries dispatched to a \modelinstance at a single point in time.} are dispatched to \modelinstances according to a provided load-balancing strategy.

\tech adds an \encoder and a \decoder on the frontend along with $\frac{m}{k}$ instances of a \paritymodel. Each \paritymodel uses the same amount of resources (e.g., compute, network) as a \basemodel. \tech thus adds $\frac{1}{k}$ resource overhead. 

As \query batches are dispatched, they are placed in a \textit{\codinggroup} consisting of $k$ batches that have been consecutively dispatched. A \codinggroup acts similarly to a ``stripe'' in erasure-coded storage systems; the \query batches of a \codinggroup are encoded to create a single ``\parityBatch.'' Encoding takes place across individual \queries of a \codinggroup: the $i$th \queries of each of the $k$ \query batches in a \codinggroup are encoded to produce the $i$th \query of the \parityBatch. Encoding \textit{does not} delay \query dispatching as \query batches are immediately handled by the load balancer when they are formed, and placed in a \codinggroup for later encoding. The \parityBatch is dispatched to a \paritymodel and the output resulting from inference over the \paritymodel is returned to the frontend. Encoding is performed on the frontend rather than on a \paritymodel so as to incur only $\frac{1}{k}$ network bandwidth overhead. Otherwise, all \queries would need to be replicated to a \paritymodel prior to encoding, which would incur $2\times$ network bandwidth overhead.

\Predictions that are returned to the frontend by \modelinstances are immediately returned to clients.\footnote{Returning \predictions from \modelinstances to the frontend is \textit{not} a new requirement imposed by \tech. This is standard in systems with a frontend, like Clipper~\cite{crankshaw2017clipper}.} \tech's \decoder is only used when any one of the $k$ \prediction batches from a \codinggroup is unavailable. The \decoder uses the outputs of the \paritymodel and the $(k-1)$ available \modelinstances to reconstruct an approximation of the unavailable \prediction batch. Approximate \predictions are returned only when \predictions from the \basemodel are unavailable, and can be annotated so that they are appropriately handled by clients.

\subsection{\Encdec design space} \label{sec:design:ec}
\tech's approach of introducing and learning a \paritymodel enables the use of simple, fast erasure codes to reconstruct unavailable \predictions. There are many \encdec designs that \tech can support, opening up a rich design space in \tech's framework. In this paper, we will illustrate the power of \tech's approach by using the dead-simple addition/subtraction erasure code described in \Section\ref{sec:background:ec}, and showing that even with the simplest choice of the \encdec, \tech significantly reduces tail latency and helps improve overall accuracy in the presence of unavailabilities.

We choose this simple encoder and decoder to showcase \tech's applicability to a variety of inference tasks including image classification, speech recognition, and object localization.
A prediction serving system that is specialized to a specific inference task could potentially benefit from \textit{designing task-specific encoders and decoders} for use within \tech. For example, for image classification tasks, an encoder could resize and concatenate image \queries for image classification. We evaluate such task-specific encoders in \Section\ref{sec:evaluation-accuracy:specific} and show that the accuracy of reconstructed predictions does increase, as expected due to the specialization.\footnote{We note that a concurrent work~\cite{narra2019collage} focusing on image classification tasks proposes a similar concatenation approach. We discuss this in \Section\ref{sec:related}.}

Under the simple addition/subtraction \encdec, the \encoder produces a parity as the summation of \queries in an \codinggroup, i.e., $\parity = \sum_{i=1}^k \data_i$. \Queries are normalized to a common size prior to encoding, and summation is performed across corresponding features of each \query (e.g., top-right pixel of each image \query). The \decoder subtracts $(k-1)$ available \predictions from the output of the \paritymodel $\outRed$ to reconstruct an unavailable \prediction. Thus, an unavailable \prediction $\func(\data_j)$ is reconstructed as $\widehat{\func(\data_j)} = \outRed - \sum_{i \neq j}^k \func(\data_i)$. 

\subsection{\Paritymodel design} \label{sec:design:pm}
\tech uses \nns for \paritymodels to learn a model that transforms parities into a form that enables decoding. Similar to \tech's encoder and decoder, there is a rich design space for potential \paritymodels.

% \noindent
\textbf{Training data.} A \paritymodel is trained prior to being deployed. The training data are the parities generated by the \encoder and the associated training labels are the transformations expected by the \decoder. For the simple \encdec described in \Section\ref{sec:design:ec}, with $k=2$, training data from \queries $\dataOne$ and $\dataTwo$ are $(\dataOne + \dataTwo)$ and labels are $(\outOne + \outTwo)$.

Training data is generated using \queries that are representative of those issued to the \basemodel for inference. A \paritymodel is trained using the same dataset used for training the \basemodel, whenever available. Thus, if the \basemodel was trained using the CIFAR-10~\cite{cifar} dataset, samples from CIFAR-10 are used as \queries $\dataOne, \ldots, \dataK$ that are encoded together to generate training samples for the \paritymodel. Labels are generated by performing inference with the \basemodel to obtain $\outOne, \ldots, \outK$ and summing these \predictions to form the desired \paritymodel output. When a labeled dataset is available, \tech can also use as labels the summation of the true labels for \queries.

If the dataset used for training the \basemodel is not available, a \paritymodel can be trained using \queries that have been issued to \tech for inference on the \basemodel. The \predictions that result from inference on the \basemodel are used to form labels for the \paritymodel. In this case, as expected, \tech can deliver benefits only after the \paritymodel has been trained to a sufficient degree. 

% \noindent
\textbf{\Nn architecture.} The design space of \nn architectures is large and presents a tradeoff between model accuracy and runtime. For example, increasing the number of layers in a \nn may lead to increased accuracy at the expense of increased runtime.

In order for a \paritymodel to help in mitigating slowdowns, the average runtime of a \paritymodel should be equal to or less than that of the \basemodel. When the \basemodel is a \nn, as is the case for state-of-the-art techniques, one way to enforce this by using the same \nn architecture for the \paritymodel as is used for the \basemodel. Thus, if the \basemodel is a ResNet-18~\cite{he2016deep} architecture, the \paritymodel can also use ResNet-18, but with parameters trained using the procedure described above, rather than for the task of the \basemodel. Note that because a \paritymodel is trained for a different task than the \basemodel, it computes a different function than the \basemodel. As a \nn's architecture determines its runtime, this approach ensures that the \paritymodel has the same average runtime as the \basemodel. 

In general, a \paritymodel is not required to use the same architecture as the \basemodel. In cases where it is necessary or preferable to use a different \nn architecture for the \paritymodel than is used for the \basemodel, such as when the \basemodel is not a \nn, one can potentially be designed via techniques like neural architecture search~\cite{zoph2016neural}. However, we do not focus on this case in this work.

\begin{figure}[t]
	\centering
    \includegraphics[width=\linewidth]{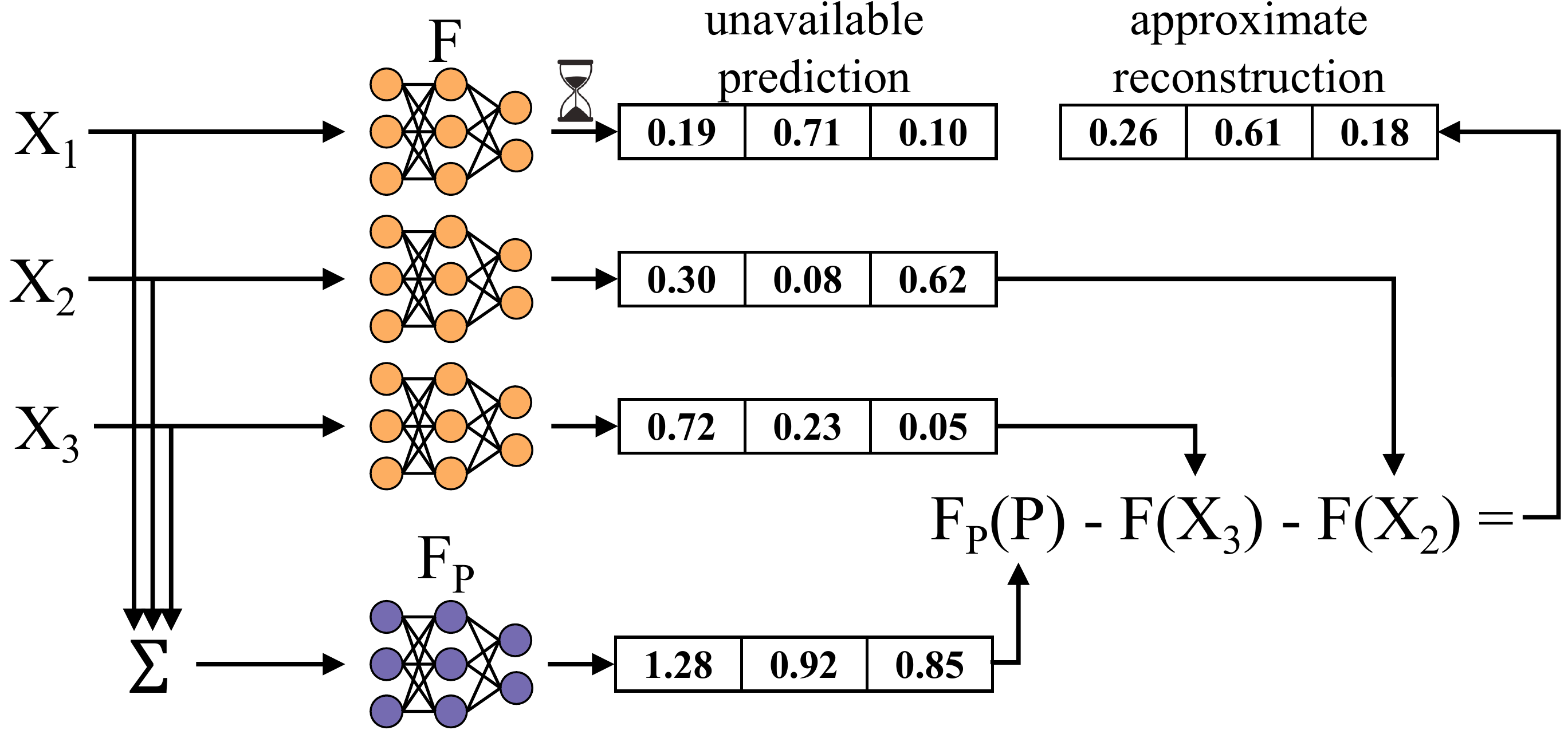}
    \caption{Example of \tech ($k=3$) mitigating a slowdown.}
    \label{fig:implementation:example}
    \vspace{-0.1in}
\end{figure}

\subsection{Example} \label{sec:design:example}
Figure~\ref{fig:implementation:example} shows an example of how \tech mitigates unavailability of any one of three \modelinstances (i.e., $k=3$). \Queries $\dataOne, \dataTwo, \dataThree$ are dispatched to three separate \modelinstances for inference on \basemodel $\func$ to return predictions $\outOne, \outTwo, \outThree$. The learning task here is classification across $n$ classes. Each $\func(\data_i)$ is thus a vector of $n$ floating-points ($n=3$ in Figure~\ref{fig:implementation:example}). 
As \queries are dispatched to \modelinstances, they are encoded ($\Sigma$) to generate a parity $\parity = (\dataOne + \dataTwo + \dataThree)$. The parity is dispatched to a \paritymodel $\funcRed$ to produce $\outRed$. In this example, the \modelinstance processing $\dataOne$ is slow. The \decoder reconstructs this unavailable \prediction as $(\outRed - \outThree - \outTwo)$. The reconstruction provides a reasonable approximation of the true \prediction that would have been returned had the \modelinstance not been slow (labeled as ``unavailable prediction'').

\subsection{Handling concurrent unavailabilities} \label{sec:design:concurrent}
\tech can accommodate concurrent unavailabilities by using \decoder{s} parameterized with $r > 1$. In this case, $r$ separate \paritymodels are trained to produce different transformations of a parity query. For example, consider having $k=2$,  $r=2$ and \queries $\dataOne$ and $\dataTwo$, with parity $\parity = (\dataOne + \dataTwo)$. One parity model is trained to transform $\parity$ into $\outOne + \outTwo$, while the second is trained to transform $\parity$ into $\outOne + 2\outTwo$. The \decoder reconstructs the initial $k$ \predictions using any $k$ out of the $(k+r)$ \predictions from \basemodels and \paritymodels.
\section{Evaluation of Accuracy}\label{sec:evaluation-accuracy}
In this section, we evaluate \tech's ability to accurately reconstruct unavailable \predictions.

\newcommand{\degradedmode}{degraded mode\xspace}
\newcommand{\Degradedmode}{Degraded mode\xspace}
\newcommand{\Degraded}{Degraded\xspace}
\newcommand{\accAvail}{A_a\xspace}
\newcommand{\accUnavail}{A_d\xspace}
\newcommand{\accOverall}{A_o\xspace}
\newcommand{\fracUnavail}{f_u\xspace}

\begin{figure*}[t]
    \centering
    \begin{minipage}[t]{0.44\textwidth}
        \centering
        \newlength\figureheight
        \newlength\figurewidth
        \setlength\figureheight{1.5in}
        \setlength\figurewidth{\textwidth}
        % This file was created by matplotlib2tikz v0.7.1.
\begin{tikzpicture}[
        hatch distance/.store in=\hatchdistance,
        hatch distance=5pt,
        hatch thickness/.store in=\hatchthickness,
        hatch thickness=0.5pt]
\pgfplotsset{label style={font=\footnotesize}, 
             tick label style={font=\footnotesize},
             legend style={font=\footnotesize},
             every non boxed x axis/.append style={x axis line style=-},
             every non boxed y axis/.append style={y axis line style=-},
             axis lines=left,
             /pgfplots/ybar legend/.style={
                /pgfplots/legend image code/.code={%
                    \draw[##1,/tikz/.cd,yshift=-0.25em]
                    (0cm,0cm) rectangle (12pt,6pt);
                },
            },
        }

% Custom hatches        
\makeatletter
    \pgfdeclarepatternformonly[\hatchdistance,\hatchthickness]{flexible hatch}
    {\pgfqpoint{0pt}{0pt}}
    {\pgfqpoint{\hatchdistance}{\hatchdistance}}
    {\pgfpoint{\hatchdistance-1pt}{\hatchdistance-1pt}}%
    {
        \pgfsetcolor{\tikz@pattern@color}
        \pgfsetlinewidth{\hatchthickness}
        \pgfpathmoveto{\pgfqpoint{0pt}{0pt}}
        \pgfpathlineto{\pgfqpoint{\hatchdistance}{\hatchdistance}}
        \pgfusepath{stroke}
    }
      
% Purple
\definecolor{color0}{rgb}{0.992156862745098,0.682352941176471,0.380392156862745}
\definecolor{color1}{rgb}{0.458823529411765,0.419607843137255,0.694117647058824}
\definecolor{color2}{rgb}{0.937254901960784,0.231372549019608,0.172549019607843}
        
% Blue        
% \definecolor{color0}{rgb}{0.992156862745098,0.682352941176471,0.380392156862745}
% \definecolor{color1}{rgb}{0.192156862745098,0.509803921568627,0.741176470588235}
% \definecolor{color2}{rgb}{0.937254901960784,0.231372549019608,0.172549019607843}
\begin{axis}[
height=\figureheight,
legend cell align={left},
legend columns=2,
legend style={at={(0.5,1.04)}, anchor=south, draw=none,
              /tikz/every even column/.append style={column sep=0.2cm}},
tick align=outside,
tick pos=left,
width=\figurewidth,
x grid style={white!69.01960784313725!black},
xlabel style={text width=\figurewidth,align=center},
xlabel={Dataset},
xmin=-0.38, xmax=5.78,
xtick={0.2,1.2,2.2,3.2,4.2,5.2},
xticklabel style={align=center},
xticklabels={MNIST,Fashion,Cat/Dog,Speech,Cifar10,Cifar100},
y grid style={gray, opacity=0.3},
ymajorgrids,
ylabel={Accuracy (Percent)},
ylabel style={at={(axis description cs:0.05,.5)}},
ymin=0, ymax=100
]
\draw[draw=black,fill=color0,thick] (axis cs:-0.1,0) rectangle (axis cs:0.1,99.2);
\addlegendimage{ybar,ybar legend,draw=black,fill=color0,thick};
\addlegendentry{Available ($\accAvail$)}

\draw[draw=black,fill=color0,thick] (axis cs:0.9,0) rectangle (axis cs:1.1,92.9);
\draw[draw=black,fill=color0,thick] (axis cs:1.9,0) rectangle (axis cs:2.1,97.1);
\draw[draw=black,fill=color0,thick] (axis cs:2.9,0) rectangle (axis cs:3.1,94.54);
\draw[draw=black,fill=color0,thick] (axis cs:3.9,0) rectangle (axis cs:4.1,93.5);
\draw[draw=black,fill=color0,thick] (axis cs:4.9,0) rectangle (axis cs:5.1,92.5);
\draw[draw=black,fill=color1,thick] (axis cs:0.1,0) rectangle (axis cs:0.3,99.21);
\addlegendimage{ybar,ybar legend,draw=black,fill=color1,thick};
\addlegendentry{\tech \Degraded ($\accUnavail$)}

\draw[draw=black,fill=color1,thick] (axis cs:1.1,0) rectangle (axis cs:1.3,91.11);
\draw[draw=black,fill=color1,thick] (axis cs:2.1,0) rectangle (axis cs:2.3,92.8);
\draw[draw=black,fill=color1,thick] (axis cs:3.1,0) rectangle (axis cs:3.3,91.48);
\draw[draw=black,fill=color1,thick] (axis cs:4.1,0) rectangle (axis cs:4.3,88.68);
\draw[draw=black,fill=color1,thick] (axis cs:5.1,0) rectangle (axis cs:5.3,86.1);
\draw[draw=black,fill=color2,thick,postaction={pattern=flexible hatch, pattern color=black}] (axis cs:0.3,0) rectangle (axis cs:0.5,10);
\addlegendimage{ybar,ybar legend,draw=black,fill=color2,thick,postaction={pattern=flexible hatch, pattern color=black}};
\addlegendentry{Default \Degraded ($\accUnavail$)}

\draw[draw=black,fill=color2,thick,postaction={pattern=flexible hatch, pattern color=black}] (axis cs:1.3,0) rectangle (axis cs:1.5,10);
\draw[draw=black,fill=color2,thick,postaction={pattern=flexible hatch, pattern color=black}] (axis cs:2.3,0) rectangle (axis cs:2.5,50);
\draw[draw=black,fill=color2,thick,postaction={pattern=flexible hatch, pattern color=black}] (axis cs:3.3,0) rectangle (axis cs:3.5,3.33);
\draw[draw=black,fill=color2,thick,postaction={pattern=flexible hatch, pattern color=black}] (axis cs:4.3,0) rectangle (axis cs:4.5,10);
\draw[draw=black,fill=color2,thick,postaction={pattern=flexible hatch, pattern color=black}]  (axis cs:5.3,0) rectangle (axis cs:5.5,5);
\end{axis}
\end{tikzpicture}
        \vspace{-0.2in}
      	\caption{Accuracies of reconstructed \predictions compared to returning a default response when \predictions from the \basemodel are unavailable ($\accUnavail$). ``Available'' is the accuracy achieved when \basemodel \predictions are available ($\accAvail$). \tech uses $k=2$ and the generic \encdec.
      	}
        \label{fig:eval-accuracy:datasets}
    \end{minipage}\hfill
    \begin{minipage}[t]{0.35\textwidth}
        \centering
        \setlength\figureheight{1.5in}
        \setlength\figurewidth{\textwidth}
        % This file was created by matplotlib2tikz v0.7.1.
\begin{tikzpicture}
\pgfplotsset{label style={font=\footnotesize}, 
             tick label style={font=\small},
             legend style={font=\footnotesize},
             every non boxed x axis/.append style={x axis line style=-},
             every non boxed y axis/.append style={y axis line style=-},
             axis lines=left,}
         
% Purple    
\definecolor{color0}{rgb}{0.458823529411765,0.419607843137255,0.694117647058824}

% Blue
% \definecolor{color0}{rgb}{0.192156862745098,0.509803921568627,0.741176470588235}
\definecolor{color1}{rgb}{0.870588235294118,0.176470588235294,0.149019607843137}
\definecolor{color2}{rgb}{0.992156862745098,0.682352941176471,0.380392156862745}
\begin{axis}[
height=\figureheight,
legend cell align={left},
legend columns=2,
legend style={at={(0.5,1.04)}, anchor=south, draw=none,
              /tikz/every even column/.append style={column sep=0.2cm}},
tick align=outside,
tick pos=left,
width=\figurewidth,
x grid style={gray, opacity=0.3},
xlabel style={text width=1.1\figurewidth,align=center},
xlabel={Fraction Unavailable ($\fracUnavail$)},
xmajorgrids,
xmin=0, xmax=0.1,
xtick={0, 0.025, 0.05, 0.075, 0.1},%,0.2,0.3,0.4},
xticklabels={0, 0.025, 0.05, 0.075, 0.1},%,0.2,0.3,0.4},
y grid style={gray, opacity=0.3},
ylabel={Accuracy (Percent)},
ylabel style={at={(axis description cs:0.07,.5)}},
ymajorgrids,
ymin=80, ymax=100
]
\addplot [ultra thick, color2, forget plot]
table {%
0 93.5
0.1 93.5
0.2 93.5
0.3 93.5
0.4 93.5
};
\addplot [ultra thick, color0]
table {%
0 93.5
0.1 93.018
0.2 92.536
0.3 92.054
0.4 91.572
};
\addlegendentry{\tech k=2}
\addplot [ultra thick, color0, dashed]
table {%
0 93.5
0.1 91.536
0.2 89.572
0.3 87.608
0.4 85.644
};
\addlegendentry{\tech k=3}
\addplot [ultra thick, color0, dash pattern=on 1pt off 3pt on 3pt off 3pt]
table {%
0 93.5
0.1 89.442
0.2 85.384
0.3 81.326
0.4 77.268
};
\addlegendentry{\tech k=4}
\addplot [ultra thick, color1, dotted]
table {%
0 93.5
0.1 85.15
0.2 76.8
0.3 68.45
0.4 60.1
};
\addlegendentry{Default}
\end{axis}
\end{tikzpicture}
        \vspace{-0.2in}
        \caption{Overall accuracy ($\accOverall$) of \predictions on CIFAR-10 as the fraction of \predictions that are unavailable ($\fracUnavail$) increases. The horizontal orange line shows the accuracy of the ResNet-18 \basemodel ($\accAvail$).}
        \label{fig:eval-accuracy:combined-acc}
    \end{minipage}\hfill
    \begin{minipage}[t]{0.18\textwidth}
        \centering
        \setlength\figureheight{1.5in}
        \setlength\figurewidth{\textwidth}
        \input{figures/accuracy/bird.tex}
        \vspace{-0.2in}
        \caption{Example of \tech's reconstruction for object localization.
        }
    \label{fig:eval-accuracy:localization}
    \end{minipage}
\end{figure*}

\subsection{Experimentation setup} \label{sec:evaluation-accuracy:setup}
We use PyTorch~\cite{pytorch} to train separate \paritymodels for each parameter $k$, dataset, and \basemodel.

%\noindent 
\textbf{Inference tasks and models.} We evaluate \tech using popular image classification (CIFAR-10 and 100~\cite{cifar}, Cat v. Dog~\cite{elson07asirra}, Fashion-MNIST~\cite{xiao2017fashion}, and MNIST~\cite{lecun1998mnist}), speech recognition (Google Commands~\cite{warden2018speech}), and object localization (CUB-200~\cite{welinder2010caltech}) tasks with varying degrees of complexity.

As described in \Section\ref{sec:design:pm}, a \paritymodel uses the same \nn architecture as the \basemodel. We consider five different architectures: a multi-layer perception (MLP),\footnote{The MLP has two hidden layers with 200 and 100 units and uses ReLU activation functions.} LeNet-5~\cite{lecun1998gradient}, VGG-11~\cite{simonyan2015very}, ResNet-18, and ResNet-152~\cite{he2016deep}. The former two are simpler \nns while the others are variants of state-of-the-art \nns.

%\noindent
\textbf{Parameters.} We consider values for parameter $k$ of 2, 3, and 4, corresponding to 33\%, 25\%, and 20\% redundancy, respectively. We use the Adam optimizer~\cite{kingma2015adam} with learning rate of 0.001, L2-regularization of $10^{-5}$, and minibatch sizes between 32 and 64. Convolutional layers are initialized by the uniform Xavier technique~\cite{glorot2010understanding}, biases are initialized to zero, and other weights are initialized from $\mathcal{N}(0, 0.01)$.

\textbf{Encoder and decoder.} Unless otherwise specified, we use the generic addition encoder and subtraction decoder described in \Section\ref{sec:design:ec}. We showcase the benefit of employing task-specific encoders and decoders within \tech in \Section\ref{sec:evaluation-accuracy:specific}.

%\noindent
\textbf{Loss function.} While there are many loss functions that could be used in training a \paritymodel, we use the mean-squared-error (MSE) between the output of the \paritymodel and the desired output as the loss function. We choose MSE rather than a task-specific loss function (e.g., cross-entropy) to make \tech applicable to many inference tasks. 

%\noindent
\textbf{Metrics.} Analyzing erasure codes for storage and communication involves reasoning about performance under normal operation (when unavailability does not occur) and performance in ``\degradedmode'' (when unavailability occurs and reconstruction is required).
These different modes of operation are similarly present for inference.
The overall accuracy of any approach is calculated based on its accuracy when \predictions from the \basemodel are available ($\accAvail$) and its accuracy when these \predictions are unavailable ($\accUnavail$, ``\degradedmode''). If $\fracUnavail$ fraction of \basemodel \predictions are unavailable, the overall accuracy ($\accOverall$) is:
\begin{equation} \label{eqn:overall-acc}
\accOverall = (1-\fracUnavail)\accAvail + \fracUnavail\accUnavail
\end{equation}
\tech aims to achieve high $\accUnavail$; it does not change the accuracy when \predictions from the \basemodel are available ($\accAvail$). We report both $\accOverall$ and $\accUnavail$.

All accuracies are evaluated using test datasets, which are not used in training. Test samples are randomly placed into groups of $k$ and encoded to produce a parity. For each parity $\parity$, we compute the output of inference on the \paritymodel as $\outRed$. During decoding, we use $\outRed$ to reconstruct $\widehat{\func(\data_i)}$ for each $\data_i$ that was used in encoding $\parity$, simulating every scenario of one \prediction being unavailable. Each $\widehat{\func(\data_i)}$ is compared to the true label for $\data_i$. For CIFAR-100, we report top-5 accuracy, as is common (i.e., the fraction for which the true class of $\data_i$ is in the top 5 of $\widehat{\func(\data_i)}$).

%\noindent
\textbf{Baseline.} We compare \tech to the approach used in Clipper~\cite{crankshaw2017clipper} to deal with unavailability: if a \prediction is not returned to the frontend by its \sla, a default \prediction is returned. This approach is motivated by observations from production services that it is better to return an incorrect \prediction than a late one~\cite{agarwal2014laser}. However, this results in \predictions that are no better than random guesses.

\subsection{Results}\label{sec:evaluation-accuracy:results}
Figure~\ref{fig:eval-accuracy:datasets} shows the accuracy of the \basemodel ($\accAvail$) along with the \degradedmode accuracy ($\accUnavail$) of \tech with $k=2$ and the default approach for image classification and speech recognition tasks using ResNet-18.\footnote{VGG-11 is used for the speech dataset and ResNet-152 for CIFAR-100.} \tech \textit{improves \degradedmode accuracy by 41-89\%} compared to returning a default \prediction. Interestingly, even when comparing \tech's \degradedmode accuracy to the \basemodel when \predictions are not slow or failed, \tech's reconstructed \predictions are at most 6.5\% less accurate than those the \basemodel would return if it were available. As Figure~\ref{fig:eval-accuracy:combined-acc} illustrates, this enables \tech to \textit{maintain high overall accuracy ($\accOverall$) in the face of unavailability}. For this example at expected levels of unavailability (i.e., $\fracUnavail$ less than 10\%), \tech's overall accuracy is at most $0.4\%$, $1.9\%$, and $4.1\%$ lower than when all \predictions are available at $k$ values of 2, 3, and 4, respectively. This indicates a tradeoff between \tech's parameter $k$, which controls resource-efficiency and resilience, and the accuracy of reconstructed \predictions, which we discuss in \Section\ref{sec:evaluation-accuracy:k}. In contrast, the overall accuracy when returning default \predictions in this setting drops by over $8.3\%$.

\begin{figure}[t]
    \centering
        \setlength\figureheight{1.5in}
        \setlength\figurewidth{\linewidth}
        % This file was created by matplotlib2tikz v0.7.1.
\begin{tikzpicture}[
        hatch distance/.store in=\hatchdistance,
        hatch distance=5pt,
        hatch thickness/.store in=\hatchthickness,
        hatch thickness=0.5pt]
\pgfplotsset{label style={font=\footnotesize}, 
             tick label style={font=\footnotesize},
             legend style={font=\footnotesize},
             every non boxed x axis/.append style={x axis line style=-},
             every non boxed y axis/.append style={y axis line style=-},
             axis lines=left,
             /pgfplots/ybar legend/.style={
                /pgfplots/legend image code/.code={%
                    \draw[##1,/tikz/.cd,yshift=-0.25em]
                    (0cm,0cm) rectangle (12pt,6pt);
                },
            },
        }
        
% Custom hatches        
% \makeatletter
%     \pgfdeclarepatternformonly[\hatchdistance,\hatchthickness]{flexible hatch}
%     {\pgfqpoint{0pt}{0pt}}
%     {\pgfqpoint{\hatchdistance}{\hatchdistance}}
%     {\pgfpoint{\hatchdistance-1pt}{\hatchdistance-1pt}}%
%     {
%         \pgfsetcolor{\tikz@pattern@color}
%         \pgfsetlinewidth{\hatchthickness}
%         \pgfpathmoveto{\pgfqpoint{0pt}{0pt}}
%         \pgfpathlineto{\pgfqpoint{\hatchdistance}{\hatchdistance}}
%         \pgfusepath{stroke}
%     }

% Purple        
\definecolor{color0}{rgb}{0.992156862745098,0.682352941176471,0.380392156862745}
\definecolor{color1}{rgb}{0.458823529411765,0.419607843137255,0.694117647058824}
\definecolor{color2}{rgb}{0.737254901960784,0.741176470588235,0.862745098039216}
\definecolor{color3}{rgb}{0.937254901960784,0.929411764705882,0.96078431372549}
\definecolor{color4}{rgb}{0.937254901960784,0.231372549019608,0.172549019607843}
% Blue
% \definecolor{color0}{rgb}{0.992156862745098,0.682352941176471,0.380392156862745}
% \definecolor{color1}{rgb}{0.192156862745098,0.509803921568627,0.741176470588235}
% \definecolor{color2}{rgb}{0.619607843137255,0.792156862745098,0.882352941176471}
% \definecolor{color3}{rgb}{0.870588235294118,0.92156862745098,0.968627450980392}
% \definecolor{color4}{rgb}{0.937254901960784,0.231372549019608,0.172549019607843}
\begin{axis}[
height=\figureheight,
legend cell align={left},
legend columns=3,
legend style={at={(0.5,1.04)}, anchor=south, draw=none,
              /tikz/every even column/.append style={column sep=0.2cm}},
tick align=outside,
tick pos=left,
width=\figurewidth,
x grid style={white!69.01960784313725!black},
xlabel style={text width=\figurewidth,align=center},
xlabel={Dataset},
xmin=-0.3625, xmax=5.9625,
xtick={0.3,1.3,2.3,3.3,4.3,5.3},
xticklabels={MNIST,Fashion,Cat/Dog,Speech,Cifar10,Cifar100},
y grid style={gray, opacity=0.3},
ymajorgrids,
ylabel={Accuracy (Percent)},
ylabel style={at={(axis description cs:0.05,.5)}},
ymin=0, ymax=100
]
\draw[draw=black,fill=color0,thick] (axis cs:-0.075,0) rectangle (axis cs:0.075,99.2);
\addlegendimage{ybar,ybar legend,draw=black,fill=color0,thick};
\addlegendentry{Available ($\accAvail$)}
\draw[draw=black,fill=color0,thick] (axis cs:0.925,0) rectangle (axis cs:1.075,92.9);
\draw[draw=black,fill=color0,thick] (axis cs:1.925,0) rectangle (axis cs:2.075,97.1);
\draw[draw=black,fill=color0,thick] (axis cs:2.925,0) rectangle (axis cs:3.075,94.54);
\draw[draw=black,fill=color0,thick] (axis cs:3.925,0) rectangle (axis cs:4.075,93.5);
\draw[draw=black,fill=color0,thick] (axis cs:4.925,0) rectangle (axis cs:5.075,92.5);
\draw[draw=black,fill=color1,thick] (axis cs:0.075,0) rectangle (axis cs:0.225,99.21);
\addlegendimage{ybar,ybar legend,draw=black,fill=color1,thick};
\addlegendentry{\tech k=2 ($\accUnavail$)}
\draw[draw=black,fill=color1,thick] (axis cs:1.075,0) rectangle (axis cs:1.225,91.11);
\draw[draw=black,fill=color1,thick] (axis cs:2.075,0) rectangle (axis cs:2.225,92.8);
\draw[draw=black,fill=color1,thick] (axis cs:3.075,0) rectangle (axis cs:3.225,91.48);
\draw[draw=black,fill=color1,thick] (axis cs:4.075,0) rectangle (axis cs:4.225,88.68);
\draw[draw=black,fill=color1,thick] (axis cs:5.075,0) rectangle (axis cs:5.225,86.1);
\draw[draw=black,fill=color2,thick] (axis cs:0.225,0) rectangle (axis cs:0.375,98.28);
\addlegendimage{ybar,ybar legend,draw=black,fill=color2,thick};
\addlegendentry{\tech k=3 ($\accUnavail$)}
\draw[draw=black,fill=color2,thick] (axis cs:1.225,0) rectangle (axis cs:1.375,87.86);
\draw[draw=black,fill=color2,thick] (axis cs:2.225,0) rectangle (axis cs:2.375,86.15);
\draw[draw=black,fill=color2,thick] (axis cs:3.225,0) rectangle (axis cs:3.375,83.92);
\draw[draw=black,fill=color2,thick] (axis cs:4.225,0) rectangle (axis cs:4.375,73.86);
\draw[draw=black,fill=color2,thick] (axis cs:5.225,0) rectangle (axis cs:5.375,69.93);
\draw[draw=black,fill=color3,thick] (axis cs:0.375,0) rectangle (axis cs:0.525,96.39);
\addlegendimage{ybar,ybar legend,draw=black,fill=color3,thick};
\addlegendentry{\tech k=4 ($\accUnavail$)}
\draw[draw=black,fill=color3,thick] (axis cs:1.375,0) rectangle (axis cs:1.525,84.21);
\draw[draw=black,fill=color3,thick] (axis cs:2.375,0) rectangle (axis cs:2.525,77.28);
\draw[draw=black,fill=color3,thick] (axis cs:3.375,0) rectangle (axis cs:3.525,75.6);
\draw[draw=black,fill=color3,thick] (axis cs:4.375,0) rectangle (axis cs:4.525,52.92);
\draw[draw=black,fill=color3,thick] (axis cs:5.375,0) rectangle (axis cs:5.525,43.94);
\draw[draw=black,fill=color4,thick,postaction={pattern=flexible hatch, pattern color=black}] (axis cs:0.525,0) rectangle (axis cs:0.675,10);
\addlegendimage{ybar,ybar legend,draw=black,fill=color4,thick,postaction={pattern=flexible hatch, pattern color=black}};
\addlegendentry{Default ($\accUnavail$)}
\draw[draw=black,fill=color4,thick,postaction={pattern=flexible hatch, pattern color=black}] (axis cs:1.525,0) rectangle (axis cs:1.675,10);
\draw[draw=black,fill=color4,thick,postaction={pattern=flexible hatch, pattern color=black}] (axis cs:2.525,0) rectangle (axis cs:2.675,50);
\draw[draw=black,fill=color4,thick,postaction={pattern=flexible hatch, pattern color=black}] (axis cs:3.525,0) rectangle (axis cs:3.675,3.33);
\draw[draw=black,fill=color4,thick,postaction={pattern=flexible hatch, pattern color=black}] (axis cs:4.525,0) rectangle (axis cs:4.675,10);
\draw[draw=black,fill=color4,thick,postaction={pattern=flexible hatch, pattern color=black}] (axis cs:5.525,0) rectangle (axis cs:5.675,5);
\end{axis}
\end{tikzpicture}
        \vspace{-0.3in}
      	\caption{Accuracies of \predictions reconstructed by \tech with $k=2,3,4$ and using the generic \encdec compared to returning a default response when \basemodel \predictions are unavailable ($\accUnavail$).
      	}
        \label{fig:eval-accuracy:k}
        \vspace{-0.1in}
\end{figure}

%\noindent
\subsubsection{Inference tasks} \label{sec:evaluation-accuracy:dataset}
\tech achieves high \degradedmode accuracy with $k=2$ for all image classification and speech recognition datasets considered. For these tasks, \degradedmode accuracy is at most 6.5\% lower than when \predictions are not slow or failed and is up to 89\% more accurate than returning a default \prediction. These improvements hold for a variety of \nn architectures. For example, on the Fashion-MNIST dataset, \tech's reconstructions for all of the MLP, LeNet-5, and ResNet-18 models are 70-81\% more accurate than returning a default \prediction.

\textbf{Object localization task.} We next evaluate \tech on object localization, which is a regression task. The goal in this task is to predict the coordinates of a bounding box surrounding an object of interest in an image. As a proof of concept of \tech's applicability for this task, we evaluate \tech on the Caltech-UCSD Birds dataset~\cite{welinder2010caltech} using ResNet-18. The performance metric for localization tasks is the intersection over union (IoU): the IoU between two bounding boxes is computed as the area of their intersection divided by the area of their union. IoU values fall between 0 and 1, with an IoU of 1 corresponding to identical boxes, and an IoU of 0 corresponding to boxes with no overlap.

Figure~\ref{fig:eval-accuracy:localization} shows the bounding boxes returned by the \basemodel and \tech's reconstruction for an example image. For this example, the \basemodel has an IoU of 0.880 and \tech's reconstruction has an IoU of 0.611. \tech's reconstruction captures the gist of the localization and would serve as a reasonable approximation in the face of unavailability. Returning a reasonable default \prediction would be infeasible for this regression task. On the entire dataset, the \basemodel achieves an average IoU of 0.945 with ground-truth bounding boxes. In \degradedmode, \tech's reconstructions with $k=2$ achieve an average IoU of 0.674.

\begin{figure}[t]
    \centering
        \includegraphics[width=0.8\linewidth]{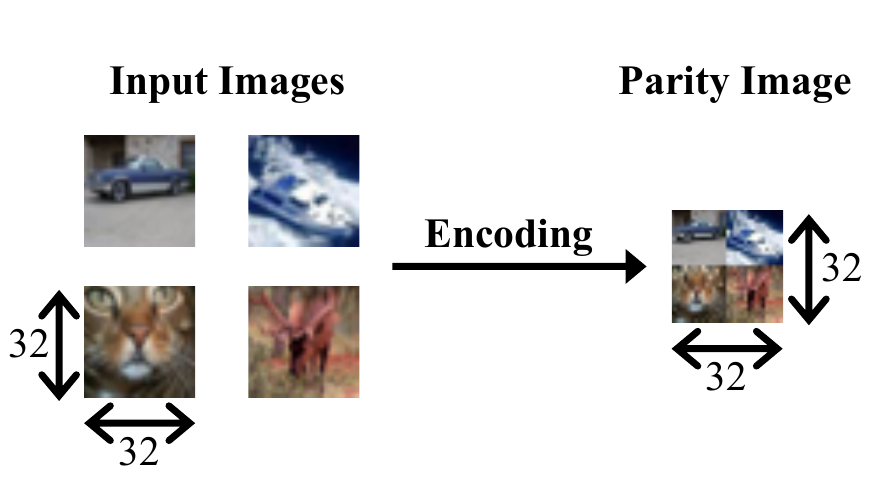}
        %\vspace{-0.3in}
      	\caption{Example of an image-classification-specific encoder with $k=4$ on the CIFAR-10 dataset.
      	}
        \label{fig:eval-accuracy:specific}
        \vspace{-0.1in}
\end{figure}

\subsubsection{Varying redundancy via parameter $\mathbf{k}$}
\label{sec:evaluation-accuracy:k}
Figure~\ref{fig:eval-accuracy:k} shows \tech's \degradedmode accuracy with $k=2,3,4$ as compared to returning a default \prediction. \tech's \degradedmode accuracy decreases with increasing parameter $k$. 
As parameter $k$ is increased, features from more \queries are packed into a single parity \query, making the parity \query noisier and making it challenging to learn a \paritymodel. 
This indicates a tradeoff between the value of parameter $k$ (i.e., redundancy) and \degradedmode accuracy. This is similar to the  performance tradeoffs that occur with increasing $k$ in the use of erasure codes in storage systems. Even at higher values of $k$, \tech's \degradedmode accuracy is still \textit{significantly higher than returning a default prediction, resulting in similar overall accuracy as the \basemodel at expected levels unavailability, as shown in Figure~\ref{fig:eval-accuracy:combined-acc}.}

\subsubsection{Inference task-specific encoders and decoders} \label{sec:evaluation-accuracy:specific}
As described in \Section\ref{sec:design:ec}, \tech's framework of introducing and learning a \paritymodel enables a large design space for possible encoders and decoders. So far, all evaluation results have used the simplest, general choice of using addition and subtraction as encoder and decoder (described in \Section\ref{sec:design:ec}), which allowed us to showcase \tech's applicability to a variety of inference tasks including image classification, speech recognition, and object localization. We now showcase the breadth of \tech's framework by evaluating \tech's accuracy when employing, alternate, encoders and decoders that are specific to a particular inference task.

We showcase the benefit of task-specific encoders and decoders for image classification tasks. We design an encoder specialized for image classification which takes in $k$ image \queries, and downsizes and concatenates them into a single parity \query. For example, as shown in Figure~\ref{fig:eval-accuracy:specific}, given $k=4$ images from the CIFAR-10 dataset (each of which have $32\times32\times3$ features), each image is resized to have $16 \times 16 \times 3$ features and concatenated together. The resultant parity \query is $2\times2$ grid of these resized images, and thus has a total of $32\times32\times3$ features, the same amount as a single image \query. We continue to use the subtraction decoder alongside this task-specific encoder.

As expected, due to the specialization to the task at hand, using this task-specific encoder leads to improved \degradedmode accuracy compared to using the general addition and subtraction code. For example, at $k$ values of 2 and 4 on the CIFAR-10 dataset, the task-specific encoder achieves a \degradedmode accuracy of 89\% and 74\%, respectively. Further, on the 1000-class ImageNet dataset (ILSVRC 2012~\cite{russakovsky2015imagenet}) with $k=2$ and using ResNet-50 models for both the \basemodel and \paritymodel, this approach achieves a 61\% top-5 \degradedmode accuracy. These results highlight the potential of using inference-task-specific encoders and decoders within \tech's framework.\footnote{We note that a concurrent work~\cite{narra2019collage} focusing on image classification tasks proposes a similar concatenation approach. We discuss this in \Section\ref{sec:related}.}

\section{Evaluation of Tail Latency Reduction} \label{sec:evaluation-system}
We next evaluate \tech's ability to reduce tail latency. The highlights of the evaluation results are as follows:
\begin{denseitemize}
	\item \tech serves predictions with predictable latencies by significantly reducing tail latency: In the presence of load imbalance, \tech reduces 99.9th percentile latency by up to $\highlightPercentDiff\%$, bringing tail latency up to $\highlightTailToMed\times$ closer to median latency, while maintaining the same median (\S\ref{sec:evaluation-system:query_rate}). Even with very little load imbalance, \tech reduces the gap between tail latency and median latency by up to $2.3\times$ (\Section\ref{sec:evaluation-system:network}). These benefits hold for different inference hardware and a wide range of query rates.
	\item \tech is effective with various batch sizes (\Section\ref{sec:evaluation-system:batch}).
    \item \tech's approach of introducing and learning \paritymodels enables using encoders and decoders with negligible latencies (less than $\SI{200}{\microsecond}$ and $\SI{20}{\microsecond}$ respectively) (\S\ref{sec:evaluation-system:components}).
    \item \tech reduces tail latency while maintaining simpler development and deployment than other hand-crafted approaches such as deploying approximate models (\S\ref{sec:evaluation-system:cheap}).
\end{denseitemize}

\subsection{Implementation and Evaluation Setup} \label{sec:evaluation-system:background}
\textbf{Implementation.} We have built \tech atop Clipper~\cite{crankshaw2017clipper}, a popular open-source \modelserver. We implement \tech's \encdec on the Clipper frontend in C++. The \basemodels and \paritymodels are PyTorch~\cite{pytorch} models in Docker containers, as is standard in Clipper. We disable the \prediction caching feature in Clipper to evaluate end-to-end latency, however we note that \tech does not preclude the use of \prediction caching. For concreteness, we showcase \tech on an image classification workload. We use OpenCV~\cite{opencv} for pixel-level \encoder operations. We use the addition encoder and subtraction decoder described in \Section\ref{sec:evaluation-accuracy:specific} in all latency evaluations.

%\noindent
\textbf{Baselines.} We consider as a baseline a \modelserver with the same number of instances as \tech but using all additional instances for deploying extra copies of the \basemodel. We call this baseline ``\baseEqual.'' For a setting of parameter $k$ on a cluster with $m$ \modelinstances for \basemodels, both \tech and \baseEqual use $\frac{m}{k}$ additional \modelinstances. \tech uses these extra instances to deploy \paritymodels, whereas the \baseEqual baseline hosts extra \basemodels on these instances.
These extra instances enable the baseline to reduce system load, which reduces tail latency and provides an apples-to-apples comparison. 
We compare \tech to another baseline, namely, deploying approximate backup models, in \Section\ref{sec:evaluation-system:cheap}.

%\noindent
\textbf{Cluster setup.} All experiments are run on Amazon EC2. We evaluate \tech on two different cluster setups to mimic various production \modelserving settings.
\begin{denseitemize}
    \item \textbf{GPU cluster.} Each \modelinstance is a \texttt{p2.xlarge} instance with one NVIDIA K80 GPU. We use 12 instances for \basemodels and $\frac{12}{k}$ additional instances for redundancy. With $k=2$ there are thus 18 instances.
    \item \textbf{High-performance CPU cluster.} Each model instance is a \texttt{c5.xlarge} instance, which Amazon recommends for  inference~\cite{aws-c5}. We use 24 instances for \basemodels and $\frac{24}{k}$ additional instances for redundancy.
    This emulates certain production \modelservices that use CPUs for inference (e.g., Facebook~\cite{hazelwood2018applied,park2018deep}, Microsoft~\cite{zhang2018deepcpu}).
    This cluster is larger than the GPU cluster since the CPU instances are less expensive than GPU instances.
\end{denseitemize}
We use a single frontend of type \texttt{c5.9xlarge}. We use this larger instance for the frontend to sustain high aggregate network bandwidth to \modelinstances (10 Gbps). Each instance uses AWS ENA networking and we observe bandwidth of 1-2 Gbps between each GPU instance and the frontend and of 4-5 Gbps between each CPU instance and the frontend.

\begin{figure*}[t]
  \centering
  \begin{minipage}[t]{0.63\textwidth}
    \begin{subfigure}[t]{0.5\textwidth}
        \centering
        \setlength\figureheight{1.7in}
        \setlength\figurewidth{\textwidth}
        % This file was created by matplotlib2tikz v0.7.1.
\begin{tikzpicture}

\pgfplotsset{label style={font=\small}, 
             tick label style={font=\small},
             legend style={font=\footnotesize},
             every non boxed x axis/.append style={x axis line style=-},
             every non boxed y axis/.append style={y axis line style=-},
             axis lines=left}

\definecolor{color0}{rgb}{0.870588235294118,0.176470588235294,0.149019607843137}

% Purple
\definecolor{color1}{rgb}{0.415686274509804,0.317647058823529,0.63921568627451}
% Blue
% \definecolor{color1}{rgb}{0.192156862745098,0.509803921568627,0.741176470588235}

\begin{axis}[
height=\figureheight,
legend cell align={left},
legend columns=2,
legend style={at={(0.5,1.)}, anchor=south, draw=none,
            /tikz/every even column/.append style={column sep=0.2cm}},
tick align=outside,
tick pos=left,
width=\figurewidth,
xtick={150, 210, 270, 330, 390},
x grid style={gray, opacity=0.3},
xlabel={Query Rate (qps)},
xmajorgrids,
xmin=138, xmax=402,
y grid style={gray, opacity=0.3},
ylabel={Latency (ms)},
ylabel style={at={(axis description cs:0.1,.5)}},
ymajorgrids,
ymin=18.4815, ymax=60
]
\path [draw=black, semithick]
(axis cs:150,20.29)
--(axis cs:150,20.64);

\path [draw=black, semithick]
(axis cs:180,20.24)
--(axis cs:180,20.38);

\path [draw=black, semithick]
(axis cs:210,20.23)
--(axis cs:210,20.29);

\path [draw=black, semithick]
(axis cs:240,20.17)
--(axis cs:240,20.39);

\path [draw=black, semithick]
(axis cs:270,20.12)
--(axis cs:270,20.52);

\path [draw=black, semithick]
(axis cs:300,20.39)
--(axis cs:300,20.45);

\path [draw=black, semithick]
(axis cs:330,20.27)
--(axis cs:330,20.34);

\path [draw=black, semithick]
(axis cs:360,20.29)
--(axis cs:360,20.46);

\path [draw=black, semithick]
(axis cs:390,20.29)
--(axis cs:390,20.44);

\path [draw=black, semithick]
(axis cs:150,53.8)
--(axis cs:150,55.56);

\path [draw=black, semithick]
(axis cs:180,51.78)
--(axis cs:180,53.49);

\path [draw=black, semithick]
(axis cs:210,53.11)
--(axis cs:210,58.75);

\path [draw=black, semithick]
(axis cs:240,51.53)
--(axis cs:240,55.08);

\path [draw=black, semithick]
(axis cs:270,51.54)
--(axis cs:270,53.98);

\path [draw=black, semithick]
(axis cs:300,49.36)
--(axis cs:300,52.05);

\path [draw=black, semithick]
(axis cs:330,48.19)
--(axis cs:330,51.66);

\path [draw=black, semithick]
(axis cs:360,48.45)
--(axis cs:360,53.59);

\path [draw=black, semithick]
(axis cs:390,49.33)
--(axis cs:390,52.35);

\path [draw=black, semithick]
(axis cs:150,20.48)
--(axis cs:150,20.81);

\path [draw=black, semithick]
(axis cs:180,20.36)
--(axis cs:180,20.52);

\path [draw=black, semithick]
(axis cs:210,20.43)
--(axis cs:210,20.57);

\path [draw=black, semithick]
(axis cs:240,20.42)
--(axis cs:240,20.52);

\path [draw=black, semithick]
(axis cs:270,20.39)
--(axis cs:270,20.67);

\path [draw=black, semithick]
(axis cs:300,20.5)
--(axis cs:300,20.62);

\path [draw=black, semithick]
(axis cs:330,20.49)
--(axis cs:330,20.64);

\path [draw=black, semithick]
(axis cs:360,20.49)
--(axis cs:360,20.78);

\path [draw=black, semithick]
(axis cs:390,20.51)
--(axis cs:390,20.76);

\path [draw=black, semithick]
(axis cs:150,33.39)
--(axis cs:150,33.63);

\path [draw=black, semithick]
(axis cs:180,31.64)
--(axis cs:180,32.73);

\path [draw=black, semithick]
(axis cs:210,31.15)
--(axis cs:210,31.94);

\path [draw=black, semithick]
(axis cs:240,30.78)
--(axis cs:240,30.93);

\path [draw=black, semithick]
(axis cs:270,30.44)
--(axis cs:270,30.8);

\path [draw=black, semithick]
(axis cs:300,29.83)
--(axis cs:300,30.38);

\path [draw=black, semithick]
(axis cs:330,29.91)
--(axis cs:330,30.5);

\path [draw=black, semithick]
(axis cs:360,29.77)
--(axis cs:360,31.55);

\path [draw=black, semithick]
(axis cs:390,30.27)
--(axis cs:390,31.2);

\addplot [semithick, black, mark=-, mark size=1.7, mark options={solid}, only marks, forget plot]
table {%
150 20.29
180 20.24
210 20.23
240 20.17
270 20.12
300 20.39
330 20.27
360 20.29
390 20.29
};
\addplot [semithick, black, mark=-, mark size=1.7, mark options={solid}, only marks, forget plot]
table {%
150 20.64
180 20.38
210 20.29
240 20.39
270 20.52
300 20.45
330 20.34
360 20.46
390 20.44
};
\addplot [semithick, black, mark=-, mark size=1.7, mark options={solid}, only marks, forget plot]
table {%
150 53.8
180 51.78
210 53.11
240 51.53
270 51.54
300 49.36
330 48.19
360 48.45
390 49.33
};
\addplot [semithick, black, mark=-, mark size=1.7, mark options={solid}, only marks, forget plot]
table {%
150 55.56
180 53.49
210 58.75
240 55.08
270 53.98
300 52.05
330 51.66
360 53.59
390 52.35
};
\addplot [semithick, black, mark=-, mark size=1.7, mark options={solid}, only marks, forget plot]
table {%
150 20.48
180 20.36
210 20.43
240 20.42
270 20.39
300 20.5
330 20.49
360 20.49
390 20.51
};
\addplot [semithick, black, mark=-, mark size=1.7, mark options={solid}, only marks, forget plot]
table {%
150 20.81
180 20.52
210 20.57
240 20.52
270 20.67
300 20.62
330 20.64
360 20.78
390 20.76
};
\addplot [semithick, black, mark=-, mark size=1.7, mark options={solid}, only marks, forget plot]
table {%
150 33.39
180 31.64
210 31.15
240 30.78
270 30.44
300 29.83
330 29.91
360 29.77
390 30.27
};
\addplot [semithick, black, mark=-, mark size=1.7, mark options={solid}, only marks, forget plot]
table {%
150 33.63
180 32.73
210 31.94
240 30.93
270 30.8
300 30.38
330 30.5
360 31.55
390 31.2
};
\addplot [semithick, color0, mark=square*, mark size=1.7, mark options={solid,fill=white}]
table {%
150 20.37
180 20.28
210 20.25
240 20.26
270 20.32
300 20.41
330 20.33
360 20.44
390 20.33
};
\addlegendentry{\baseEqualShort med.}
\addplot [semithick, color1, mark=*, mark size=1.7, mark options={solid}]
table {%
150 20.52
180 20.51
210 20.45
240 20.45
270 20.46
300 20.57
330 20.56
360 20.67
390 20.64
};
\addlegendentry{\tech med.}
\addplot [semithick, color0, dashed, mark=square*, mark size=1.7, mark options={solid,fill=white}]
table {%
150 55.41
180 52.62
210 53.31
240 53.65
270 52.84
300 51.2
330 51.37
360 49.25
390 49.74
};
\addlegendentry{\baseEqualShort 99.9th}
\addplot [semithick, color1, dashed, mark=*, mark size=1.7, mark options={solid}]
table {%
150 33.42
180 32.57
210 31.23
240 30.79
270 30.78
300 30.09
330 30.08
360 30.37
390 30.64
};
\addlegendentry{\tech 99.9th}
\end{axis}

\end{tikzpicture}
       	\caption{GPU cluster}
        \label{fig:rate-p2}
      \end{subfigure} \hfill
      \begin{subfigure}[t]{0.5\textwidth}
        \centering
        \setlength\figureheight{1.7in}
        \setlength\figurewidth{\textwidth}
        % This file was created by matplotlib2tikz v0.7.1.
\begin{tikzpicture}

\pgfplotsset{label style={font=\small}, 
             tick label style={font=\small},
             legend style={font=\footnotesize},
             every non boxed x axis/.append style={x axis line style=-},
             every non boxed y axis/.append style={y axis line style=-},
             axis lines=left}

\definecolor{color0}{rgb}{0.870588235294118,0.176470588235294,0.149019607843137}

% Purple
\definecolor{color1}{rgb}{0.415686274509804,0.317647058823529,0.63921568627451}
% Blue
% \definecolor{color1}{rgb}{0.192156862745098,0.509803921568627,0.741176470588235}

\begin{axis}[
height=\figureheight,
legend cell align={left},
legend columns=2,
legend style={at={(0.5,1.)}, anchor=south, draw=none,
            /tikz/every even column/.append style={column sep=0.2cm}},
tick align=outside,
tick pos=left,
width=\figurewidth,
xtick={120,144,168,192,216,240},
x grid style={gray, opacity=0.3},
xlabel={Query Rate (qps)},
xmajorgrids,
xmin=114, xmax=246,
y grid style={gray, opacity=0.3},
ytick={75, 125, 175, 225},
ylabel={Latency (ms)},
ylabel style={at={(axis description cs:0.05,.5)}},
ymajorgrids,
ymin=63.799, ymax=230.361
]
\path [draw=black, semithick]
(axis cs:120,71.53)
--(axis cs:120,71.69);

\path [draw=black, semithick]
(axis cs:144,71.42)
--(axis cs:144,71.62);

\path [draw=black, semithick]
(axis cs:168,71.5)
--(axis cs:168,71.63);

\path [draw=black, semithick]
(axis cs:192,71.59)
--(axis cs:192,71.73);

\path [draw=black, semithick]
(axis cs:216,71.43)
--(axis cs:216,71.5);

\path [draw=black, semithick]
(axis cs:240,71.37)
--(axis cs:240,71.48);

\path [draw=black, semithick]
(axis cs:120,210.67)
--(axis cs:120,219.94);

\path [draw=black, semithick]
(axis cs:144,211.83)
--(axis cs:144,220.88);

\path [draw=black, semithick]
(axis cs:168,213.39)
--(axis cs:168,216.26);

\path [draw=black, semithick]
(axis cs:192,211.31)
--(axis cs:192,222.79);

\path [draw=black, semithick]
(axis cs:216,209.16)
--(axis cs:216,219.53);

\path [draw=black, semithick]
(axis cs:240,203.36)
--(axis cs:240,208.81);

\path [draw=black, semithick]
(axis cs:120,72.18)
--(axis cs:120,72.49);

\path [draw=black, semithick]
(axis cs:144,71.99)
--(axis cs:144,72.21);

\path [draw=black, semithick]
(axis cs:168,72.6)
--(axis cs:168,72.74);

\path [draw=black, semithick]
(axis cs:192,72.48)
--(axis cs:192,72.74);

\path [draw=black, semithick]
(axis cs:216,72.34)
--(axis cs:216,72.8);

\path [draw=black, semithick]
(axis cs:240,72.74)
--(axis cs:240,72.8);

\path [draw=black, semithick]
(axis cs:120,116.13)
--(axis cs:120,119.77);

\path [draw=black, semithick]
(axis cs:144,115.49)
--(axis cs:144,116.62);

\path [draw=black, semithick]
(axis cs:168,115.52)
--(axis cs:168,116.3);

\path [draw=black, semithick]
(axis cs:192,114.2)
--(axis cs:192,116.07);

\path [draw=black, semithick]
(axis cs:216,114.3)
--(axis cs:216,117.5);

\path [draw=black, semithick]
(axis cs:240,117.32)
--(axis cs:240,118.74);

\addplot [semithick, black, mark=-, mark size=1.7, mark options={solid}, only marks, forget plot]
table {%
120 71.53
144 71.42
168 71.5
192 71.59
216 71.43
240 71.37
};
\addplot [semithick, black, mark=-, mark size=1.7, mark options={solid}, only marks, forget plot]
table {%
120 71.69
144 71.62
168 71.63
192 71.73
216 71.5
240 71.48
};
\addplot [semithick, black, mark=-, mark size=1.7, mark options={solid}, only marks, forget plot]
table {%
120 210.67
144 211.83
168 213.39
192 211.31
216 209.16
240 203.36
};
\addplot [semithick, black, mark=-, mark size=1.7, mark options={solid}, only marks, forget plot]
table {%
120 219.94
144 220.88
168 216.26
192 222.79
216 219.53
240 208.81
};
\addplot [semithick, black, mark=-, mark size=1.7, mark options={solid}, only marks, forget plot]
table {%
120 72.18
144 71.99
168 72.6
192 72.48
216 72.34
240 72.74
};
\addplot [semithick, black, mark=-, mark size=1.7, mark options={solid}, only marks, forget plot]
table {%
120 72.49
144 72.21
168 72.74
192 72.74
216 72.8
240 72.8
};
\addplot [semithick, black, mark=-, mark size=1.7, mark options={solid}, only marks, forget plot]
table {%
120 116.13
144 115.49
168 115.52
192 114.2
216 114.3
240 117.32
};
\addplot [semithick, black, mark=-, mark size=1.7, mark options={solid}, only marks, forget plot]
table {%
120 119.77
144 116.62
168 116.3
192 116.07
216 117.5
240 118.74
};
\addplot [semithick, color0, mark=square*, mark size=1.7, mark options={solid,fill=white}]
table {%
120 71.58
144 71.55
168 71.54
192 71.68
216 71.49
240 71.48
};
\addlegendentry{\baseEqualShort med.}
\addplot [semithick, color1, mark=*, mark size=1.7, mark options={solid}]
table {%
120 72.31
144 72.17
168 72.69
192 72.48
216 72.5
240 72.75
};
\addlegendentry{\tech med.}
\addplot [semithick, color0, dashed, mark=square*, mark size=1.7, mark options={solid,fill=white}]
table {%
120 219.9
144 220.13
168 213.59
192 221.48
216 213.63
240 207.97
};
\addlegendentry{\baseEqualShort 99.9th}
\addplot [semithick, color1, dashed, mark=*, mark size=1.7, mark options={solid}]
table {%
120 118.7
144 115.72
168 116.05
192 114.6
216 115.04
240 117.46
};
\addlegendentry{\tech 99.9th}
\end{axis}

\end{tikzpicture}
        \vspace{-0.15in}
          \caption{CPU cluster}
          \label{fig:rate-c5}
      \end{subfigure}
      \vspace{-0.1in}
      \caption{Latencies of \tech and \baseEqual (\baseEqualShort). The CPU cluster has twice as many instances as the GPU cluster and thus sustains comparable load.
      }
      \label{fig:eval-systems:k_qr}
  \end{minipage}\hfill
  \begin{minipage}[t]{0.35\textwidth}
    \centering
    \setlength\figureheight{1.7in}
    \setlength\figurewidth{\textwidth}
    % This file was created by matplotlib2tikz v0.7.1.
\begin{tikzpicture}

\pgfplotsset{label style={font=\small}, 
             tick label style={font=\small},
             legend style={font=\footnotesize},
             every non boxed x axis/.append style={x axis line style=-},
             every non boxed y axis/.append style={y axis line style=-},
             axis lines=left,
             /pgfplots/ybar legend/.style={
                /pgfplots/legend image code/.code={%
                    \draw[##1,/tikz/.cd,yshift=-0.25em]
                    (0cm,0cm) rectangle (12pt,6pt);
                },
            },
        }

% Purple
\definecolor{color0}{rgb}{0.415686274509804,0.317647058823529,0.63921568627451}
\definecolor{color1}{rgb}{0.619607843137255,0.603921568627451,0.784313725490196}
\definecolor{color2}{rgb}{0.796078431372549,0.788235294117647,0.886274509803922}
\definecolor{color3}{rgb}{0.949019607843137,0.941176470588235,0.968627450980392}

% Blue
% \definecolor{color0}{rgb}{0.129411764705882,0.443137254901961,0.709803921568627}
% \definecolor{color1}{rgb}{0.419607843137255,0.682352941176471,0.83921568627451}
% \definecolor{color2}{rgb}{0.741176470588235,0.843137254901961,0.905882352941176}
% \definecolor{color3}{rgb}{0.937254901960784,0.952941176470588,1}

\begin{axis}[
height=\figureheight,
legend cell align={left},
legend columns=2,
legend style={at={(0.5,1.)}, anchor=south, draw=none,
            /tikz/every even column/.append style={column sep=0.2cm}},
tick align=outside,
tick pos=left,
width=\figurewidth,
x grid style={white!69.01960784313725!black},
xlabel={Metric},
xmin=-0.305, xmax=4.755,
xtick={0.225,1.225,2.225,3.225,4.225},
xticklabels={Med.,Mean,99th,99.5th,99.9th},
%y grid style={white!69.01960784313725!black},
ylabel={Latency (ms)},
ylabel style={at={(axis description cs:0.1,.5)}},
ytick={0, 10, 20, 30, 40, 50},
ymin=0, ymax=55.482,
y grid style={gray, opacity=0.3},
ymajorgrids
]
\draw[draw=black,fill=color0,thick] (axis cs:-0.075,0) rectangle (axis cs:0.075,20.46);
\addlegendimage{ybar,ybar legend,draw=black,fill=color0,thick};
\addlegendentry{\tech k=2 (33\%)}

\draw[draw=black,fill=color0,thick] (axis cs:0.925,0) rectangle (axis cs:1.075,19.71);
\draw[draw=black,fill=color0,thick] (axis cs:1.925,0) rectangle (axis cs:2.075,25.39);
\draw[draw=black,fill=color0,thick] (axis cs:2.925,0) rectangle (axis cs:3.075,26.82);
\draw[draw=black,fill=color0,thick] (axis cs:3.925,0) rectangle (axis cs:4.075,30.8);
\draw[draw=black,fill=color1,thick] (axis cs:0.075,0) rectangle (axis cs:0.225,20.65);
\addlegendimage{ybar,ybar legend,draw=black,fill=color1,thick};
\addlegendentry{\tech k=3 (25\%)}

\draw[draw=black,fill=color1,thick] (axis cs:1.075,0) rectangle (axis cs:1.225,19.93);
\draw[draw=black,fill=color1,thick] (axis cs:2.075,0) rectangle (axis cs:2.225,26.93);
\draw[draw=black,fill=color1,thick] (axis cs:3.075,0) rectangle (axis cs:3.225,28.7);
\draw[draw=black,fill=color1,thick] (axis cs:4.075,0) rectangle (axis cs:4.225,33.27);
\draw[draw=black,fill=color2,thick] (axis cs:0.225,0) rectangle (axis cs:0.375,20.58);
\addlegendimage{ybar,ybar legend,draw=black,fill=color2,thick};
\addlegendentry{\tech k=4 (20\%)}

\draw[draw=black,fill=color2,thick] (axis cs:1.225,0) rectangle (axis cs:1.375,19.96);
\draw[draw=black,fill=color2,thick] (axis cs:2.225,0) rectangle (axis cs:2.375,28.05);
\draw[draw=black,fill=color2,thick] (axis cs:3.225,0) rectangle (axis cs:3.375,29.92);
\draw[draw=black,fill=color2,thick] (axis cs:4.225,0) rectangle (axis cs:4.375,34.69);
\draw[draw=black,fill=color3,thick] (axis cs:0.375,0) rectangle (axis cs:0.525,20.32);
\addlegendimage{ybar,ybar legend,draw=black,fill=color3,thick};
\addlegendentry{\baseEqual (33\%)}

\draw[draw=black,fill=color3,thick] (axis cs:1.375,0) rectangle (axis cs:1.525,20.24);
\draw[draw=black,fill=color3,thick] (axis cs:2.375,0) rectangle (axis cs:2.525,33.3);
\draw[draw=black,fill=color3,thick] (axis cs:3.375,0) rectangle (axis cs:3.525,38.29);
\draw[draw=black,fill=color3,thick] (axis cs:4.375,0) rectangle (axis cs:4.525,52.84);
\path [draw=black, thick]
(axis cs:0,20.46)
--(axis cs:0,20.46);

\path [draw=black, thick]
(axis cs:1,19.71)
--(axis cs:1,19.71);

\path [draw=black, thick]
(axis cs:2,25.39)
--(axis cs:2,25.39);

\path [draw=black, thick]
(axis cs:3,26.82)
--(axis cs:3,26.82);

\path [draw=black, thick]
(axis cs:4,30.8)
--(axis cs:4,30.8);

\path [draw=black, thick]
(axis cs:0.15,20.65)
--(axis cs:0.15,20.65);

\path [draw=black, thick]
(axis cs:1.15,19.93)
--(axis cs:1.15,19.93);

\path [draw=black, thick]
(axis cs:2.15,26.93)
--(axis cs:2.15,26.93);

\path [draw=black, thick]
(axis cs:3.15,28.7)
--(axis cs:3.15,28.7);

\path [draw=black, thick]
(axis cs:4.15,33.27)
--(axis cs:4.15,33.27);

\path [draw=black, thick]
(axis cs:0.3,20.58)
--(axis cs:0.3,20.58);

\path [draw=black, thick]
(axis cs:1.3,19.96)
--(axis cs:1.3,19.96);

\path [draw=black, thick]
(axis cs:2.3,28.05)
--(axis cs:2.3,28.05);

\path [draw=black, thick]
(axis cs:3.3,29.92)
--(axis cs:3.3,29.92);

\path [draw=black, thick]
(axis cs:4.3,34.69)
--(axis cs:4.3,34.69);

\path [draw=black, thick]
(axis cs:0.45,20.32)
--(axis cs:0.45,20.32);

\path [draw=black, thick]
(axis cs:1.45,20.24)
--(axis cs:1.45,20.24);

\path [draw=black, thick]
(axis cs:2.45,33.3)
--(axis cs:2.45,33.3);

\path [draw=black, thick]
(axis cs:3.45,38.29)
--(axis cs:3.45,38.29);

\path [draw=black, thick]
(axis cs:4.45,52.84)
--(axis cs:4.45,52.84);

\end{axis}

\end{tikzpicture}
    \vspace{-0.35in}
    \caption{Latencies of \tech at varying values of $k$ compared to the strongest baseline. The amount of redundancy used in each configuration is listed in parentheses.
    }
    \label{fig:eval-system:k}
  \end{minipage}
\end{figure*}

%\noindent
\textbf{Queries and \basemodels.} Recall that accuracy results were presented for various tasks and \basemodels in \Section\ref{sec:evaluation-accuracy}. For latency evaluations we choose one of these models, ResNet-18~\cite{he2016deep}. We use ResNet-18 rather than a larger model like ResNet-152, which would have a longer runtime, to provide a more challenging scenario in which \tech must reconstruct \predictions with low latency. \Queries are drawn from the Cat v. Dog~\cite{elson07asirra} dataset. These higher-resolution images test the ability of \tech's \encoder to operate with low latency.\footnote{
While CIFAR-10/100 are more difficult tasks for training a model than Cat v. Dog, their low resolution makes them computationally inexpensive. This makes Cat v. Dog a more realistic workload for evaluating latency.} We modify \basemodels and \paritymodels to return vectors of 1000 floating points as \predictions to create a more computationally challenging decoding scenario in which there are 1000 classes in each \prediction, rather than the usual 2 classes for this task.

Client instances send 100-thousand \queries to the frontend using a variety of Poisson arrival rates. Unless otherwise noted, all experiments are run with batch size of one, as this is the preferred batch size for low latency~\cite{chung2018serving,zhang2018deepcpu}. We evaluate \tech with larger batch sizes in \Section\ref{sec:evaluation-system:batch}.

%\noindent
\textbf{Load balancing.} Both \tech and the baseline use a single-queue load balancing strategy for dispatching \queries to \modelinstances as is used in Clipper, and is optimal in reducing average response time~\cite{harchol2013performance}. The frontend maintains a single queue to which all \queries are added. \Modelinstances pull \queries from this queue when they are available. Similarly, \tech adds parity \queries to a single queue which \paritymodels pull from. Evaluation on other, sub-optimal, load balancing strategies (e.g., round-robin) revealed results that are even more favorable for \tech than those showcased below.

%\noindent
\textbf{Background traffic.} 
As it is difficult to mimic tail latency inflation scenarios without access to production systems, we evaluate \tech with various types and degrees of background load. The main form of background load we use emulates network traffic typical of data analytics workloads. Specifically, two \modelinstances are chosen at random to transfer data to one another of size randomly drawn between 128 MB and 256 MB. Unless otherwise mentioned, four of these shuffles take place concurrently. In this setting \textit{only} the cluster network is imbalanced and we do not introduce any computational multitenancy. We experiment with light multitenant computation and varying the number of shuffles in \Section\ref{sec:evaluation-system:network}.

%\noindent
\textbf{Latency metric.} All latencies measure the time between when the frontend receives a \query and when the corresponding \prediction is returned to the frontend (from a \basemodel or reconstructed). The latency of communication between clients and the frontend is not included, as this latency is not controlled by \tech. We report the median of three runs of each configuration (each with 100-thousand \queries), with error bars showing the minimum and maximum.

\newcommand{\gpuTailToMed}{2.6-3.2}
\newcommand{\cpuTailToMed}{3-3.5}
\subsection{Results} \label{sec:evaluation-system:results}
We now report \tech's reduction of tail latency.

\newcommand{\gpuQrTurningPoint}{330\xspace}
\newcommand{\gpuHighThresholdImp}{41}
\subsubsection{Varying \query rate} \label{sec:evaluation-system:query_rate}
Figure~\ref{fig:eval-systems:k_qr} shows median and 99.9th percentile latencies with $k=2$ (i.e., both \tech and the \baseEqual baseline have 33\% redundancy) on the GPU and CPU clusters. We consider query rates up until a point in which a \modelserver with no redundancy (i.e., using only $m$ instances) experiences tail latency dominated by queueing. Beyond this point, \tech could be used alongside a number of techniques that reduce queueing delays~\cite{crankshaw2018inferline}.

\tech reduces the gap between 99.9th percentile latency and median latency by \gpuTailToMed$\times$ compared to \baseEqual on the GPU cluster, and by \cpuTailToMed$\times$ on the CPU cluster. \tech's 99.9th percentile latencies are thus 38-43\% lower on the GPU cluster and 44-48\% lower on the CPU cluster. 
This enables \tech to return \predictions with more predictable latencies. As expected from any redundancy-based approach, \tech adds additional system load by issuing redundant \queries, leading to a slight increase in median latency (less than half of a millisecond, which is negligible).

\begin{figure*}[t]
  \centering
  \begin{minipage}[t]{0.32\textwidth}
        \centering
        \setlength\figureheight{1.5in}
        \setlength\figurewidth{\linewidth}
        % This file was created by matplotlib2tikz v0.7.1.
\begin{tikzpicture}

\pgfplotsset{label style={font=\small}, 
             tick label style={font=\small},
             legend style={font=\footnotesize},
             every non boxed x axis/.append style={x axis line style=-},
             every non boxed y axis/.append style={y axis line style=-},
             axis lines=left,}

\definecolor{color0}{rgb}{0.870588235294118,0.176470588235294,0.149019607843137}
% Purple
\definecolor{color1}{rgb}{0.415686274509804,0.317647058823529,0.63921568627451}
% Blue
% \definecolor{color1}{rgb}{0.192156862745098,0.509803921568627,0.741176470588235}

\begin{axis}[
height=\figureheight,
legend cell align={left},
legend columns=2,
legend style={at={(0.5,1.04)}, anchor=south, draw=none,
              /tikz/every even column/.append style={column sep=0.2cm}},
tick align=outside,
tick pos=left,
width=\figurewidth,
x grid style={gray, opacity=0.3},
xlabel={Number of background shuffles},
xmajorgrids,
xtick={2,3,4,5},
xmin=1.85, xmax=5.15,
y grid style={gray, opacity=0.3},
ylabel style={at={(axis description cs:0.05,.5)}},
ylabel={Latency (ms)},
ytick={20,30,40,50,60},
ymajorgrids,
ymin=17.9925, ymax=63.6975
]
\path [draw=black, semithick]
(axis cs:2,20.07)
--(axis cs:2,20.09);

\path [draw=black, semithick]
(axis cs:3,20.17)
--(axis cs:3,20.22);

\path [draw=black, semithick]
% (axis cs:4,20.54)
% --(axis cs:4,20.66);
(axis cs:4,20.12)
--(axis cs:4,20.52);

\path [draw=black, semithick]
(axis cs:5,20.65)
--(axis cs:5,20.7);

\path [draw=black, semithick]
(axis cs:2,42.29)
--(axis cs:2,43.53);

\path [draw=black, semithick]
(axis cs:3,46.41)
--(axis cs:3,47.97);

\path [draw=black, semithick]
% (axis cs:4,50.45)
% --(axis cs:4,55.32);
(axis cs:4,51.54)
--(axis cs:4,53.98);

\path [draw=black, semithick]
(axis cs:5,53.33)
--(axis cs:5,61.62);

\path [draw=black, semithick]
(axis cs:2,20.27)
--(axis cs:2,20.28);

\path [draw=black, semithick]
(axis cs:3,20.33)
--(axis cs:3,20.4);

\path [draw=black, semithick]
% (axis cs:4,20.72)
% --(axis cs:4,20.82);
(axis cs:4,20.39)
--(axis cs:4,20.67);

\path [draw=black, semithick]
(axis cs:5,20.83)
--(axis cs:5,20.93);

\path [draw=black, semithick]
(axis cs:2,27.85)
--(axis cs:2,28.19);

\path [draw=black, semithick]
(axis cs:3,29.13)
--(axis cs:3,29.31);

\path [draw=black, semithick]
% (axis cs:4,30.04)
% --(axis cs:4,30.87);
(axis cs:4,30.44)
--(axis cs:4,30.8);

\path [draw=black, semithick]
(axis cs:5,31.65)
--(axis cs:5,31.87);

\addplot [semithick, black, mark=-, mark size=1.7, mark options={solid}, only marks, forget plot]
table {%
2 20.07
3 20.17
4 20.54
5 20.65
};
\addplot [semithick, black, mark=-, mark size=1.7, mark options={solid}, only marks, forget plot]
table {%
2 20.09
3 20.22
4 20.66
5 20.7
};
\addplot [semithick, black, mark=-, mark size=1.7, mark options={solid}, only marks, forget plot]
table {%
2 42.29
3 46.41
4 51.54
5 53.33
};
\addplot [semithick, black, mark=-, mark size=1.7, mark options={solid}, only marks, forget plot]
table {%
2 43.53
3 47.97
4 53.98
5 61.62
};
\addplot [semithick, black, mark=-, mark size=1.7, mark options={solid}, only marks, forget plot]
table {%
2 20.27
3 20.33
4 20.39
5 20.83
};
\addplot [semithick, black, mark=-, mark size=1.7, mark options={solid}, only marks, forget plot]
table {%
2 20.28
3 20.4
4 20.67
5 20.93
};
\addplot [semithick, black, mark=-, mark size=1.7, mark options={solid}, only marks, forget plot]
table {%
2 27.85
3 29.13
4 30.44
5 31.65
};
\addplot [semithick, black, mark=-, mark size=1.7, mark options={solid}, only marks, forget plot]
table {%
2 28.19
3 29.31
4 30.8
5 31.87
};
\addplot [semithick, color0, mark=square*, mark size=1.7, mark options={solid,fill=white}]
table {%
2 20.09
3 20.19
4 20.32
5 20.67
};
\addlegendentry{\baseEqualShort med.}
\addplot [semithick, color1, mark=*, mark size=1.7, mark options={solid}]
table {%
2 20.27
3 20.38
4 20.46
5 20.88
};
\addlegendentry{\tech med.}
\addplot [semithick, color0, dashed, mark=square*, mark size=1.7, mark options={solid,fill=white}]
table {%
2 43.06
3 47.56
4 52.84
5 59.19
};
\addlegendentry{\baseEqualShort 99.9th}
\addplot [semithick, color1, dashed, mark=*, mark size=1.7, mark options={solid}]
table {%
2 28.13
3 29.16
4 30.78
5 31.69
};
\addlegendentry{\tech 99.9th}
\end{axis}

\end{tikzpicture}
        \vspace{-0.25in}
        \caption{\tech and \baseEqual (\baseEqualShort) with varying network imbalance.}
        \label{fig:eval-system:network}
  \end{minipage}\hfill
  \begin{minipage}[t]{0.32\textwidth}
        \centering
        \setlength\figureheight{1.5in}
        \setlength\figurewidth{\linewidth}
        % This file was created by matplotlib2tikz v0.7.1.
\begin{tikzpicture}
\newcommand{\ylabelshift}{0.05}
\pgfplotsset{label style={font=\small}, 
             tick label style={font=\small},
             legend style={font=\footnotesize},
             every non boxed x axis/.append style={x axis line style=-},
             every non boxed y axis/.append style={y axis line style=-},
             axis lines=left}

\definecolor{color0}{rgb}{0.870588235294118,0.176470588235294,0.149019607843137}

% Purple
\definecolor{color1}{rgb}{0.415686274509804,0.317647058823529,0.63921568627451}
% Blue
% \definecolor{color1}{rgb}{0.192156862745098,0.509803921568627,0.741176470588235}

\begin{axis}[
height=\figureheight,
legend cell align={left},
legend columns=2,
legend style={at={(0.5,1.04)}, anchor=south, draw=none,
            /tikz/every even column/.append style={column sep=0.2cm}},
tick align=outside,
tick pos=left,
width=\figurewidth,
xtick={150, 210, 270, 330, 390},
x grid style={gray, opacity=0.3},
xlabel={Query Rate (qps)},
xmajorgrids,
xmin=138, xmax=402,
y grid style={gray, opacity=0.3},
ylabel={Latency (ms)},
ylabel style={at={(axis description cs:\ylabelshift,.5)}},
ymajorgrids,
ymin=18.973, ymax=36.287
]
\path [draw=black, semithick]
(axis cs:150,19.91)
--(axis cs:150,19.97);

\path [draw=black, semithick]
(axis cs:180,19.87)
--(axis cs:180,19.9);

\path [draw=black, semithick]
(axis cs:210,19.83)
--(axis cs:210,19.93);

\path [draw=black, semithick]
(axis cs:240,19.82)
--(axis cs:240,19.84);

\path [draw=black, semithick]
(axis cs:270,19.78)
--(axis cs:270,19.83);

\path [draw=black, semithick]
(axis cs:300,19.76)
--(axis cs:300,20.33);

\path [draw=black, semithick]
(axis cs:330,20.14)
--(axis cs:330,20.18);

\path [draw=black, semithick]
(axis cs:360,20.17)
--(axis cs:360,20.21);

\path [draw=black, semithick]
(axis cs:390,19.97)
--(axis cs:390,20.2);

\path [draw=black, semithick]
(axis cs:150,34.28)
--(axis cs:150,34.86);

\path [draw=black, semithick]
(axis cs:180,34.17)
--(axis cs:180,35.37);

\path [draw=black, semithick]
(axis cs:210,33.83)
--(axis cs:210,34.43);

\path [draw=black, semithick]
(axis cs:240,33.94)
--(axis cs:240,34.62);

\path [draw=black, semithick]
(axis cs:270,34.25)
--(axis cs:270,34.63);

\path [draw=black, semithick]
(axis cs:300,33.71)
--(axis cs:300,34.79);

\path [draw=black, semithick]
(axis cs:330,34.84)
--(axis cs:330,35.41);

\path [draw=black, semithick]
(axis cs:360,34.81)
--(axis cs:360,35.29);

\path [draw=black, semithick]
(axis cs:390,34.49)
--(axis cs:390,35.5);

\path [draw=black, semithick]
(axis cs:150,20.02)
--(axis cs:150,20.06);

\path [draw=black, semithick]
(axis cs:180,20.01)
--(axis cs:180,20.02);

\path [draw=black, semithick]
(axis cs:210,20.02)
--(axis cs:210,20.04);

\path [draw=black, semithick]
(axis cs:240,19.94)
--(axis cs:240,19.98);

\path [draw=black, semithick]
(axis cs:270,19.9)
--(axis cs:270,19.93);

\path [draw=black, semithick]
(axis cs:300,19.93)
--(axis cs:300,20.44);

\path [draw=black, semithick]
(axis cs:330,20.25)
--(axis cs:330,20.33);

\path [draw=black, semithick]
(axis cs:360,20.29)
--(axis cs:360,20.34);

\path [draw=black, semithick]
(axis cs:390,19.99)
--(axis cs:390,20.31);

\path [draw=black, semithick]
(axis cs:150,27.02)
--(axis cs:150,27.91);

\path [draw=black, semithick]
(axis cs:180,27.22)
--(axis cs:180,27.47);

\path [draw=black, semithick]
(axis cs:210,26.89)
--(axis cs:210,27.29);

\path [draw=black, semithick]
(axis cs:240,26.4)
--(axis cs:240,26.7);

\path [draw=black, semithick]
(axis cs:270,26.11)
--(axis cs:270,26.72);

\path [draw=black, semithick]
(axis cs:300,26.31)
--(axis cs:300,26.63);

\path [draw=black, semithick]
(axis cs:330,26.55)
--(axis cs:330,27.41);

\path [draw=black, semithick]
(axis cs:360,26.98)
--(axis cs:360,27.99);

\path [draw=black, semithick]
(axis cs:390,27.43)
--(axis cs:390,28.15);

\addplot [semithick, black, mark=-, mark size=1.7, mark options={solid}, only marks, forget plot]
table {%
150 19.91
180 19.87
210 19.83
240 19.82
270 19.78
300 19.76
330 20.14
360 20.17
390 19.97
};
\addplot [semithick, black, mark=-, mark size=1.7, mark options={solid}, only marks, forget plot]
table {%
150 19.97
180 19.9
210 19.93
240 19.84
270 19.83
300 20.33
330 20.18
360 20.21
390 20.2
};
\addplot [semithick, black, mark=-, mark size=1.7, mark options={solid}, only marks, forget plot]
table {%
150 34.28
180 34.17
210 33.83
240 33.94
270 34.25
300 33.71
330 34.84
360 34.81
390 34.49
};
\addplot [semithick, black, mark=-, mark size=1.7, mark options={solid}, only marks, forget plot]
table {%
150 34.86
180 35.37
210 34.43
240 34.62
270 34.63
300 34.79
330 35.41
360 35.29
390 35.5
};
\addplot [semithick, black, mark=-, mark size=1.7, mark options={solid}, only marks, forget plot]
table {%
150 20.02
180 20.01
210 20.02
240 19.94
270 19.9
300 19.93
330 20.25
360 20.29
390 19.99
};
\addplot [semithick, black, mark=-, mark size=1.7, mark options={solid}, only marks, forget plot]
table {%
150 20.06
180 20.02
210 20.04
240 19.98
270 19.93
300 20.44
330 20.33
360 20.34
390 20.31
};
\addplot [semithick, black, mark=-, mark size=1.7, mark options={solid}, only marks, forget plot]
table {%
150 27.02
180 27.22
210 26.89
240 26.4
270 26.11
300 26.31
330 26.55
360 26.98
390 27.43
};
\addplot [semithick, black, mark=-, mark size=1.7, mark options={solid}, only marks, forget plot]
table {%
150 27.91
180 27.47
210 27.29
240 26.7
270 26.72
300 26.63
330 27.41
360 27.99
390 28.15
};
\addplot [semithick, color0, mark=square*, mark size=1.7, mark options={solid,fill=white}]
table {%
150 19.92
180 19.89
210 19.9
240 19.83
270 19.81
300 19.89
330 20.18
360 20.2
390 20.04
};
\addlegendentry{\baseEqualShort med.}
\addplot [semithick, color1, mark=*, mark size=1.7, mark options={solid}]
table {%
150 20.05
180 20.02
210 20.03
240 19.97
270 19.93
300 20.36
330 20.32
360 20.29
390 20.06
};
\addlegendentry{\tech med.}
\addplot [semithick, color0, dashed, mark=square*, mark size=1.7, mark options={solid,fill=white}]
table {%
150 34.79
180 34.49
210 34.02
240 34.09
270 34.32
300 34.68
330 35.23
360 35.15
390 34.85
};
\addlegendentry{\baseEqualShort 99.9th}
\addplot [semithick, color1, dashed, mark=*, mark size=1.7, mark options={solid}]
table {%
150 27.89
180 27.42
210 26.96
240 26.48
270 26.2
300 26.55
330 26.98
360 27.61
390 27.9
};
\addlegendentry{\tech 99.9th}
\end{axis}

\end{tikzpicture}
        \vspace{-0.25in}
        \caption{\tech and \baseEqual (\baseEqualShort) with light inference multitenancy.}
        \label{fig:eval-system:multitenancy}
  \end{minipage}\hfill
  \begin{minipage}[t]{0.32\textwidth}
        \centering
        \setlength\figureheight{1.5in}
        \setlength\figurewidth{\linewidth}
        % This file was created by matplotlib2tikz v0.7.1.
\begin{tikzpicture}

\pgfplotsset{label style={font=\small}, 
             tick label style={font=\small},
             legend style={font=\footnotesize},
             every non boxed x axis/.append style={x axis line style=-},
             every non boxed y axis/.append style={y axis line style=-},
             axis lines=left,}
             
\definecolor{color0}{rgb}{0.870588235294118,0.176470588235294,0.149019607843137}
% Purple
\definecolor{color1}{rgb}{0.415686274509804,0.317647058823529,0.63921568627451}
% Blue
% \definecolor{color1}{rgb}{0.192156862745098,0.509803921568627,0.741176470588235}

\begin{axis}[
height=\figureheight,
legend cell align={left},
legend columns=2,
legend style={at={(0.5,1.04)}, anchor=south, draw=none,
              /tikz/every even column/.append style={column sep=0.2cm}},
tick align=outside,
tick pos=left,
width=\figurewidth,
x grid style={gray, opacity=0.3},
xlabel={Query Rate (qps)},
xmajorgrids,
xmin=138, xmax=402,
xtick={150, 210, 270, 330, 390},
y grid style={gray, opacity=0.3},
ylabel={Latency (ms)},
ylabel style={at={(axis description cs:0.05,.5)}},
ymajorgrids,
ymin=19.472, ymax=40
]
\path [draw=black, semithick]
(axis cs:150,20.32)
--(axis cs:150,20.54);

\path [draw=black, semithick]
(axis cs:180,20.2)
--(axis cs:180,20.5);

\path [draw=black, semithick]
(axis cs:210,20.2)
--(axis cs:210,20.5);

\path [draw=black, semithick]
(axis cs:240,20.11)
--(axis cs:240,20.49);

\path [draw=black, semithick]
(axis cs:270,20.13)
--(axis cs:270,20.5);

\path [draw=black, semithick]
(axis cs:300,20.26)
--(axis cs:300,20.43);

\path [draw=black, semithick]
(axis cs:330,20.31)
--(axis cs:330,20.44);

\path [draw=black, semithick]
(axis cs:360,20.4)
--(axis cs:360,20.43);

\path [draw=black, semithick]
(axis cs:390,20.4)
--(axis cs:390,20.47);

\path [draw=black, semithick]
(axis cs:150,27.68)
--(axis cs:150,28.38);

\path [draw=black, semithick]
(axis cs:180,28.12)
--(axis cs:180,28.61);

\path [draw=black, semithick]
(axis cs:210,29.52)
--(axis cs:210,29.89);

\path [draw=black, semithick]
(axis cs:240,30.58)
--(axis cs:240,31.92);

\path [draw=black, semithick]
(axis cs:270,33.22)
--(axis cs:270,33.7);

\path [draw=black, semithick]
(axis cs:300,34.06)
--(axis cs:300,34.51);

\path [draw=black, semithick]
(axis cs:330,35.4)
--(axis cs:330,37.36);

\path [draw=black, semithick]
(axis cs:360,36.21)
--(axis cs:360,36.75);

\path [draw=black, semithick]
(axis cs:390,38.1)
--(axis cs:390,38.21);

\path [draw=black, semithick]
(axis cs:150,20.48)
--(axis cs:150,20.81);

\path [draw=black, semithick]
(axis cs:180,20.36)
--(axis cs:180,20.52);

\path [draw=black, semithick]
(axis cs:210,20.43)
--(axis cs:210,20.57);

\path [draw=black, semithick]
(axis cs:240,20.42)
--(axis cs:240,20.52);

\path [draw=black, semithick]
(axis cs:270,20.39)
--(axis cs:270,20.67);

\path [draw=black, semithick]
(axis cs:300,20.5)
--(axis cs:300,20.62);

\path [draw=black, semithick]
(axis cs:330,20.49)
--(axis cs:330,20.64);

\path [draw=black, semithick]
(axis cs:360,20.49)
--(axis cs:360,20.78);

\path [draw=black, semithick]
(axis cs:390,20.51)
--(axis cs:390,20.76);

\path [draw=black, semithick]
(axis cs:150,33.39)
--(axis cs:150,33.63);

\path [draw=black, semithick]
(axis cs:180,31.64)
--(axis cs:180,32.73);

\path [draw=black, semithick]
(axis cs:210,31.15)
--(axis cs:210,31.94);

\path [draw=black, semithick]
(axis cs:240,30.78)
--(axis cs:240,30.93);

\path [draw=black, semithick]
(axis cs:270,30.44)
--(axis cs:270,30.8);

\path [draw=black, semithick]
(axis cs:300,29.83)
--(axis cs:300,30.38);

\path [draw=black, semithick]
(axis cs:330,29.91)
--(axis cs:330,30.5);

\path [draw=black, semithick]
(axis cs:360,29.77)
--(axis cs:360,31.55);

\path [draw=black, semithick]
(axis cs:390,30.27)
--(axis cs:390,31.2);

\addplot [semithick, black, mark=-, mark size=1.7, mark options={solid}, only marks, forget plot]
table {%
150 20.32
180 20.2
210 20.2
240 20.11
270 20.13
300 20.26
330 20.31
360 20.4
390 20.4
};
\addplot [semithick, black, mark=-, mark size=1.7, mark options={solid}, only marks, forget plot]
table {%
150 20.54
180 20.5
210 20.5
240 20.49
270 20.5
300 20.43
330 20.44
360 20.43
390 20.47
};
\addplot [semithick, black, mark=-, mark size=1.7, mark options={solid}, only marks, forget plot]
table {%
150 27.68
180 28.12
210 29.52
240 30.58
270 33.22
300 34.06
330 35.4
360 36.21
390 38.1
};
\addplot [semithick, black, mark=-, mark size=1.7, mark options={solid}, only marks, forget plot]
table {%
150 28.38
180 28.61
210 29.89
240 31.92
270 33.7
300 34.51
330 37.36
360 36.75
390 38.21
};
\addplot [semithick, black, mark=-, mark size=1.7, mark options={solid}, only marks, forget plot]
table {%
150 20.48
180 20.36
210 20.43
240 20.42
270 20.39
300 20.5
330 20.49
360 20.49
390 20.51
};
\addplot [semithick, black, mark=-, mark size=1.7, mark options={solid}, only marks, forget plot]
table {%
150 20.81
180 20.52
210 20.57
240 20.52
270 20.67
300 20.62
330 20.64
360 20.78
390 20.76
};
\addplot [semithick, black, mark=-, mark size=1.7, mark options={solid}, only marks, forget plot]
table {%
150 33.39
180 31.64
210 31.15
240 30.78
270 30.44
300 29.83
330 29.91
360 29.77
390 30.27
};
\addplot [semithick, black, mark=-, mark size=1.7, mark options={solid}, only marks, forget plot]
table {%
150 33.63
180 32.73
210 31.94
240 30.93
270 30.8
300 30.38
330 30.5
360 31.55
390 31.2
};
\addplot [semithick, color0, mark=square*, mark size=1.7, mark options={solid,fill=white}]
table {%
150 20.52
180 20.26
210 20.23
240 20.26
270 20.2
300 20.27
330 20.42
360 20.42
390 20.43
};
\addlegendentry{A.B. med.}
\addplot [semithick, color1, mark=*, mark size=1.7, mark options={solid}]
table {%
150 20.52
180 20.51
210 20.45
240 20.45
270 20.46
300 20.57
330 20.56
360 20.67
390 20.64
};
\addlegendentry{\tech med.}
\addplot [semithick, color0, dashed, mark=square*, mark size=1.7, mark options={solid,fill=white}]
table {%
150 27.93
180 28.42
210 29.69
240 30.69
270 33.66
300 34.32
330 36.15
360 36.32
390 38.12
};
\addlegendentry{A.B. 99.9th}
\addplot [semithick, color1, dashed, mark=*, mark size=1.7, mark options={solid}]
table {%
150 33.42
180 32.57
210 31.23
240 30.79
270 30.78
300 30.09
330 30.08
360 30.37
390 30.64
};
\addlegendentry{\tech 99.9th}
\end{axis}

\end{tikzpicture}
        \vspace{-0.25in}
        \caption{Latencies of \tech and using approximate backup models (A.B.).}
        \label{fig:eval-systems:approx}
  \end{minipage}
\end{figure*}

\newcommand{\gpuVaryKTailToMed}{2.5}

\subsubsection{Varying redundancy via parameter $\mathbf{k}$} \label{sec:evaluation-system:param_k}
Figure~\ref{fig:eval-system:k} shows the latencies achieved by \tech with $k$ being 2, 3, and 4, when operating at 270 qps on the GPU cluster. As $k$ increases, \tech's tail latency also increases. This is due to two factors. First, at higher values of $k$, \tech is more vulnerable to multiple \predictions in a \codinggroup being unavailable, as the \decoder requires $k-1$ \predictions from the \basemodel to be available (in addition to the output of the \paritymodel). Second, increasing $k$ increases the amount of time \tech needs to wait for $k$ \queries to arrive before encoding into a parity query. This increases the latency of the end-to-end path of reconstructing an unavailable \prediction. 

Despite these factors, \tech still \textit{reduces tail latency over the baseline that uses more resources than \tech}. At $k$ values of 3 and 4, which have 25\% and 20\% redundancy respectively, \tech reduces the difference between 99.9th percentile and median latency by up to \gpuVaryKTailToMed$\times$ compared to when \baseEqual has 33\% redundancy.

\newcommand{\batchSizeOneImp}{3.3}
\newcommand{\batchSizeTwoImp}{3.1}
\newcommand{\batchSizeFourImp}{4}

\subsubsection{Varying batch size}\label{sec:evaluation-system:batch}
Due to the low latencies required by user-facing applications, many \modelservers perform no or minimal \query batching~\cite{chung2018serving,zhang2018deepcpu, hazelwood2018applied}. For completeness, we evaluate \tech when \queries are batched for inference on the GPU cluster. \tech uses $k=2$ in these experiments and \query rate is set to 460 qps and 584 qps for batch sizes of 2 and 4, respectively. These \query rates are obtained by scaling from 300 qps used at batch size 1 based on the throughput improvement observed with increasing batch sizes.

\textit{\tech outperforms \baseEqual at all batch sizes}: at batch sizes of 2 and 4, \tech reduces 99.9th percentile latency by 43\% and 47\%. This reduces gap between 99.9th percentile and median latency by up to $\batchSizeFourImp\times$ over \baseEqual.

\subsubsection{Varying degrees and types of load imbalance} \label{sec:evaluation-system:network}
All experiments so far were run with background network imbalance, as described in \Section\ref{sec:evaluation-system:background}. \tech reduces tail latency even with lighter background network load: Figure~\ref{fig:eval-system:network} shows that when 2 and 3 concurrent background shuffles take place (as opposed to the 4 used for most experiments), \tech reduces 99.9th percentile latency over \baseEqual by 35\% and 39\%, respectively on the GPU cluster with \query rate of 270 qps. \tech's benefits increase with higher load imbalance, as \tech reduces the gap between 99.9th and median latency by $3.5\times$ over \baseEqual with 5 background shuffles. 

To evaluate \tech's resilience to a different, lighter form of load imbalance, we run light background inference tasks on \modelinstances. Specifically, we deploy ResNet-18 models on one ninth of instances using a separate copy of Clipper, and send an average query rate of less than 5\% of what the cluster can maintain. We do not add network imbalance in this setting. Figure~\ref{fig:eval-system:multitenancy} shows latencies at $k=2$ on the GPU cluster with varying \query rate. Even with this light form of imbalance, \tech reduces the gap between 99.9th percentile and median latency by up to $2.3\times$ over \baseEqual.

\subsubsection{Latency of \tech's components} \label{sec:evaluation-system:components}
\tech's latency of reconstructing unavailable \predictions consists of three components: encoding, \paritymodel inference, and decoding. \tech has median encoding latencies of $\SI{93}{\microsecond}$, $\SI{153}{\microsecond}$, and $\SI{193}{\microsecond}$, and median decoding latencies of $\SI{8}{\microsecond}$, $\SI{14}{\microsecond}$, and $\SI{19}{\microsecond}$ for $k$ values of 2, 3, and 4, respectively. As the latency of \paritymodel inference is tens of milliseconds, \textit{\tech's encoding and decoding make up a very small fraction of end-to-end reconstruction latency}. These fast encoders and decoders are enabled by \tech's new approach of introducing \paritymodels that allows it to use simple erasure codes.

\newcommand{\approxTailVariationLower}{36}
\subsubsection{Comparison to approximate backup models} \label{sec:evaluation-system:cheap}
An alternative to \tech is to replace \tech's \paritymodels with less computationally expensive models that approximate the \predictions of the \basemodel, and to replicate \queries to these approximate models. While potentially capable of returning approximate \predictions in the face of unavailability, this approach has a number of drawbacks: (1) it is unstable at expected \query rates, (2) it is inflexible to changes in hardware, limiting deployment flexibility, and (3) it requires $2\times$ network bandwidth. To showcase these drawbacks, we compare \tech (with $k=2$) to the aforementioned alternative using $\frac{m}{k}$ extra \modelinstances for approximate models. We use  MobileNet-V2~\cite{sandler2018mobilenetv2} (width factor of 0.25) as the approximate models because this model has similar accuracy (87.6\%) as \tech's reconstructions (87.4\%) for CIFAR-10.

Figure~\ref{fig:eval-systems:approx} shows the latencies of these approaches on the GPU cluster with varying \query rate. While \tech's 99.9th percentile latency varies only modestly, using approximate models results in tail latency variations of over \approxTailVariationLower\%. This variance occurs because all \queries are replicated to approximate models even though there are only $\frac{1}{k}$ as many approximate models as there are \basemodels. 
Thus, approximate models must be $k$-times faster than the \basemodel for this system to be stable. The approximate model in this case is not $k$-times faster than the \basemodel, leading to inflated tail latency due to queueing as \query rate increases.

Even if one crafted an approximate model satisfying the runtime requirement described above, the model may not be appropriate for different hardware. We find that the speedup achieved by the approximate model over the \basemodel varies substantially across different inference hardware. For example, the MobileNet-V2 approximate model is $1.4\times$ faster than the ResNet-18 \basemodel on the CPU cluster, but only $1.15\times$ faster on the GPU cluster. Thus, an approximate model designed for one hardware setup may not provide benefits on other hardware, limiting deployment flexibility. Designing an approximate model for every possible hardware setup would require iterative effort from data scientists. 

Finally, this approach uses $2\times$ network bandwidth by replicating \queries. This can be problematic, as limited bandwidth has been shown to hinder \modelserving~\cite{hauswald2015djinn,crankshaw2017clipper}.

\tech does not have any of these drawbacks. As described in \Section\ref{sec:design}, \tech's \paritymodels can be chosen to have the same average runtime as \basemodels, as showcased in the evaluation. Furthermore, \tech encodes $k$ \queries into one parity query prior to dispatching to a \paritymodel. The $\frac{m}{k}$ \paritymodels therefore receive $\frac{1}{k}$ the \query rate of the $m$ \basemodels, and thus naturally keep pace. This reduced \query rate also means that \tech adds only minor network bandwidth overhead. Further, when using the same model architecture for \paritymodels as is used for \basemodels, \tech does not face hardware-related deployment issues.
\section{Related Work}\label{sec:related}

%\noindent
\textbf{Mitigating slowdowns.}
Many approaches alleviate specific causes of slowdown. Examples of such techniques include configuration selection~\cite{venkataraman2016ernest,alipourfard2017cherrypick,yadwadkar2017selecting,li2018metis}, tenant isolation~\cite{xu2013bobtail,grosvenor2015queues,mace20162dfq,iorgulescu2018perfiso}, replica selection~\cite{suresh2015c3,hao2017mittos}, queueing disciplines~\cite{liang2013fast,joshi2014delay, shah2016redundant,gardner2015reducing}, and autoscaling~\cite{gujarati2017swayam,crankshaw2018inferline}. As these techniques are applicable only to certain types of slowdowns and are often not straightforward to weave together, they are unable to mitigate all slowdowns. In contrast, \tech is agnostic to the cause of slowdown. In \Section\ref{sec:background:bottlenecks}, we described two existing agnostic approaches to mitigating unavailability and their downsides, which \tech overcomes.

There are many techniques~\cite{recht2011hogwild,ho2013more,wei2015managed,harlap2016addressing} for mitigating slowdowns that occur in \textit{training} a model. These techniques exploit iterative computations specific to training and are thus inapplicable to alleviating slowdowns during \textit{inference}.

%\noindent
\textbf{Coded-computation.}
Most prior work related to coded-computation was discussed in detail in \Section\ref{sec:intro} and \Section\ref{sec:background}. We discuss a some of these related works in more detail below.

A recently proposed class of codes~\cite{yu2018lagrange} supports polynomial (non-linear) functions, but requires as many or more resources than replication-based approaches. Another approach~\cite{dutta2018unified} performs coded-computation over the linear operations of \nns and decodes before each non-linear operation. This requires splitting the operations of a model onto multiple servers and many decoding steps, which increases latency even when predictions are not slow or failed. In contrast, \tech uses 2-4$\times$ less resource overhead than replication and does not require neural network operations to be performed on separate servers and does not induce latency when there is no unavailability.

In a concurrent work focusing on image classification tasks, Narra \etal\cite{narra2019collage} propose to concatenate multiple images into a single image for inference on a specialized, multi-object detection model. This approach fits within \tech's framework as an example of employing task-specific encoders and decoders as discussed in \Section\ref{sec:design:ec}. In fact, the encoding proposed in~\cite{narra2019collage} is very similar to the concatenation-based image-classification-specific encoder that we proposed in \Section\ref{sec:evaluation-accuracy:specific}. In contrast, \tech is a \textit{general} framework for coding-based resilient inference, which we have shown to be \textit{applicable for a variety of inference tasks including image classification, speech recognition, and object localization}. \tech's approach of using parity models opens up a rich design space for designing encoders, decoders, and \paritymodels, enabling specialization to specific inference tasks when needed. Further, the approach proposed in~\cite{narra2019collage} results in concatenated images that can be up to $k$-times larger than the original images (e.g., for CIFAR-10 when images are not resized), which can consume up to $2\times$ additional network bandwidth. In contrast, \tech incurs only $\frac{1}{k}$ network bandwidth overhead.

%\noindent
\textbf{High performance inference methods.}
There are many techniques for reducing the average latency~\cite{viola2004robust,jacob2017quantization,wang2018idk,chen2018tvm,lee2018pretzel,nvidia-tensorrt} and improving the throughput~\cite{jiang2018mainstream,lee2018pretzel} of inference. In contrast \tech is designed for mitigating latency spikes that occur throughout \modelserving, including those that occur during model inference, network transfer, and system faults. These techniques are complementary to \tech.

%\noindent
\textbf{Accuracy-latency tradeoff.} A number of systems~\cite{zhang2017live,wang2018rafiki} and machine learning techniques~\cite{huang2017multi,hu2017anytime} \textit{actively} trade \prediction accuracy with latency. This enables these techniques to handle \query rate variation in a resource-efficient manner, but may result in lower accuracy. In contrast, \tech does not proactively degrade \prediction accuracy. Rather, any inaccuracy that comes from \tech is incurred only when a \prediction experiences slowdown or failure. 
\section{Conclusion}\label{sec:conclusion}
We have presented \tech, a general framework for imparting coding-based resilience against slowdowns and failures that occur in \modelservers. \tech overcomes the challenges of prior coding-based resilience techniques by introducing \textit{\paritymodels}  as a new building block that enable simple, fast encoders and decoders to reconstruct unavailable \predictions for a variety of inference tasks, including image classification, speech recognition, and object localization. We have built \tech atop a popular open-source \modelserver and extensively evaluated the ability of \tech's framework to reduce tail latency and improve overall accuracy under unavailability for a wide variety of inference tasks. \tech reduces the gap between 99.9th percentile and median latency by up to $\highlightTailToMed\times$ compared to approaches that use the same amount of resources, while maintaining the same median. 

\tech's framework presents a fundamentally new coding-based approach for imparting resilience to general inference tasks. Our evaluation results showcase the promise of \tech's approach for adding resource-efficient resilience to \modelservers. 

{\footnotesize \bibliographystyle{acm}
\bibliography{references}}

\end{document}